\documentclass[12pt]{article}
\usepackage{titling}
\usepackage{authblk}
\usepackage[english]{babel}

\usepackage[a4paper,top=2cm,bottom=2cm,left=3cm,right=3cm,marginparwidth=1.75cm]{geometry}

\usepackage{amsmath}
\usepackage{graphicx}
\usepackage[colorlinks=true, allcolors=blue]{hyperref}
\usepackage[font={small}]{caption}
\usepackage{subcaption}
\usepackage[labelformat=simple]{subcaption}

\usepackage[export]{adjustbox}
\usepackage{bbm}
\usepackage{comment} 
\usepackage{appendix}
\usepackage{amsmath,amssymb}
\usepackage{cancel}
\usepackage[table]{xcolor}
\definecolor{my_orange}{HTML}{FF7F00}
\definecolor{my_purple}{HTML}{984EA3}
\definecolor{my_blue}{HTML}{377EB8}
\definecolor{my_red}{HTML}{E41A1C}
\definecolor{my_green}{HTML}{4DAF4A}
\usepackage{diagbox}
\usepackage{natbib}
\usepackage{mathtools}
\mathtoolsset{showonlyrefs=true}
\setcitestyle{authoryear,aysep{()}}

\usepackage{setspace}
\title{Multiplex Dirichlet stochastic block model for clustering multidimensional compositional networks}

\author[1,2]{Iuliia Promskaia}
\author[1,2]{Adrian O'Hagan}
\author[2]{Michael Fop}
\affil[1]{Insight Research Ireland Centre for Data Analytics, Dublin, Ireland}
\affil[2]{School of Mathematics and Statistics, University College Dublin, Dublin, Ireland}
\date{}
\date{\small\href{mailto: iuliia.promskaia@insight-centre.org}{iuliia.promskaia@insight-centre.org}} 
\begin{document}
\maketitle
\begin{abstract}

Network data often represent multiple types of relations, which can also denote exchanged quantities, and are typically encompassed in a weighted multiplex. Such data frequently exhibit clustering structures, however, traditional clustering methods are not well-suited for multiplex networks. Additionally, standard methods treat edge weights in their raw form, potentially biasing clustering towards a node’s total weight capacity rather than reflecting cluster-related interaction patterns. To address this, we propose transforming edge weights into a compositional format, enabling the analysis of connection strengths in relative terms and removing the impact of nodes’ total weights. We introduce a multiplex Dirichlet stochastic block model designed for multiplex networks with compositional layers. This model accounts for sparse compositional networks and enables joint clustering across different types of interactions. We validate the model through a simulation study and apply it to the international export data from the Food and Agriculture Organization of the United Nations.

\end{abstract}

\vspace{5mm}


\section{Introduction}

Networks are often used to describe interactions or relationships between agents in a system. Network data emerge in diverse contexts, such as social networks \citep{Karrer-Newman2011,sbm_interaction_lengths,po_dc_dynamic_sbm}, transportation \citep{weighted_sbm,Aicher_weighted_sbm}, international trade \citep{Melnykov_edge_clustering, Domenico_fao_data}, financial networks \citep{Financial_networks,weighted_multilayer_sbm} and biological networks \citep{strata_multilayer_sbm,Domenico_fao_data}. As a result, statistical network analysis has become an active and growing area of research.

One common task in statistical network analysis is clustering, which aims to identify meaningful groups of units that share similar connectivity patterns or behaviours in the network. Model-based clustering \citep{Fraley_Raftery2002} is based on the idea that the data arise from a mixture of a finite number of probability distributions, with each component corresponding to a cluster \citep{bouveyron_2019}. A popular family of clustering models for network data is stochastic block models \citep[SBM,][]{SBM_review}. Initially developed to identify groups of nodes with similar connectivity patterns in binary networks \citep{Holland83,WangWong87,Wasserman87}, SBMs have since been extended to accommodate the increasing complexity and diversity of network data. For instance, to model networks describing weighted interactions, SBMs for integer weighted networks have been proposed in \cite{binomial_sbm}, \cite{Karrer-Newman2011} and \cite{Zanghi_2010}, and continuous weights have been considered in \cite{Aicher_2013,Aicher_weighted_sbm,Ludkin_2020} and \cite{weighted_sbm}. 

In many applications, agents within a system can interact in multiple distinct ways. For example, in social networks, two individuals can be connected due to family ties, living in the same neighbourhood, working for the same company, sharing the same hobby, or due to a combination of reasons. Networks that involve the same set of agents connected through multiple types of relationships are often referred to as multiplex networks \citep{social_net_anal}, with each relationship type represented as a separate layer of information. Most of the existing clustering methods have been developed for uniplex networks, i.e. networks with a single layer of data. These methods typically only allow analysis of either a specific subset of interactions or a simplified version of the multiplex network, where multiple layers are combined into a single layer. Such approaches may lead to information loss and may reduce the robustness of the analysis, as the structural characteristics of individual layers can vary noticeably.

Recent literature has increasingly focused on modelling and clustering networks with multiple layers. To address the presence of multiple layers in binary networks, \cite{multiplex_sbm_barbillon} proposed an extension of the binary SBM for the analysis of a network of researchers, with one layer represents whether researchers have sought advice from one another, while another layer captures resource transfers between their labs. A general stochastic block modelling framework for multi-layer networks, which includes dynamic and multiplex networks as special cases, is proposed in \cite{multilayer_sbm_estimation}, in which two estimation procedures based on spectral clustering and maximum likelihood are described. In \cite{strata_multilayer_sbm}, the authors use a two-step clustering procedure whereby in the first step the network layers are assigned to strata, i.e. groups of layers in which the nodes share the same connectivity and cluster assignment patterns, and then, conditional on the strata assignment, the nodes are allocated to clusters using a stochastic block model approach. In \cite{restricted_multilayer_sbm}, a version of a multilayer stochastic block model is proposed, with each individual layer exhibiting a different connectivity pattern. For the purpose of clustering in neuroimaging, a random effects stochastic block model is developed in \cite{random_effect_sbm}, 
which introduces putative mean community structures to represent group-specific behaviors, enabling each layer of the multiplex network to exhibit its own clustering structure, which deviates from the mean structure through transition probabilities. A notable work based on the ideas presented in \cite{Aicher_weighted_sbm} that performs clustering on weighted multilayer networks is \cite{weighted_multilayer_sbm}. In \cite{Melnykov_edge_clustering}, a general framework for performing model-based clustering on weighted multilayer networks is presented, with a particular focus on continuous interaction weights.

In the works above, the multiplex networks are assumed to be either binary or attributed with edge weights in their raw form. However, in some applications, the nodes' capacities to send and receive edge weights can differ significantly, potentially influencing the clustering results. A standard approach to address variability in nodes' capacities to exchange binary edges is through the degree-corrected stochastic block model \citep[DC-SBM,][]{Karrer-Newman2011}, which uses node-specific parameters to account for node heterogeneous degree distributions. This model adjusts for the differences in the number of edges the individual nodes can exchange. However, in some contexts, nodes may also differ in the amount of weight they can send or receive, an aspect distinct from their capacity based solely on edge counts. Failing to account for the total weights sent and received by individual nodes may lead to clustering solutions that reflect patterns driven by the overall weights of the sending and receiving nodes, rather than the specific weights connecting them. Using as an example the international trade networks that are examined in Section \ref{Application}, factors such as country's population and GDP have a strong influence on the country's import and export volumes and significantly less or no influence on the number of trade partners the country has. Analysing total import and export values between pairs of countries in absolute terms may reflect differences in population sizes, GDPs, and related factors. This type of pattern in international trade network analysis has been observed in \cite{Melnykov_edge_clustering}, where the authors implement model-based clustering of the European trade network, noting that the identified clustering structure and the value of the goods traded are linked to the countries' GDP.

In \cite{dir_sbm}, the authors address this issue by representing the network in a compositional format, expressing weighted relations using relative rather than absolute weights. This approach involves calculating the proportions of flow along each edge relative to a sending (or receiving) node, resulting in edge weights that sum to one for each node. To cluster such compositional networks, the authors propose the Dirichlet stochastic block model (DirSBM), which models the compositional edge weights as a mixture of Dirichlet distributions. In this work, building on the framework of \cite{dir_sbm}, we develop two key extensions to the existing Dirichlet stochastic block model. First, we extend the model to account for absent edges, enabling modeling of sparse compositional networks, hence addressing the limitation of the original DirSBM, which assumes fully connected networks, a restrictive assumption in many applications. Second, we adapt the DirSBM to multiplex networks, enabling the model to account for multiple types of interactions when clustering is performed.

The structure of this paper is as follows: we begin by introducing the Food and Agriculture Organization (FAO) data published by \cite{Domenico_fao_data} to motivate the proposed model extensions and describe the pre-processing undertaken for the analysis in Section \ref{FAO_data}. The original Dirichlet stochastic block model (DirSBM) is briefly described in Section \ref{Dirichlet_SBM}. Section \ref{Extensions_zeros} introduces the first extension of the model, which enables to model directly the presence and absence of edges. Section \ref{Extensions_layers} further extends the model to multiplex networks. Model selection using an integrated completed likelihood criterion and approximated Bayesian information criterion is addressed in Section \ref{model_selection}. The results of simulation studies assessing the clustering performance, parameter estimation and model selection of the extended model are presented in Section \ref{Sim_studies}. The proposed model is then applied to the FAO trade data \citep{Domenico_fao_data}, and the results are presented in Section \ref{Application}. The paper concludes with a discussion in Section \ref{Discussion}.

\section{FAO trade data} \label{FAO_data}

The Food and Agriculture Organization (FAO) is an agency within the United Nations, founded in 1945, with its primary focus fighting hunger and improving nutrition and food security. FAO provides support to governments and development agencies in matters related to agriculture, fisheries, forestry, and land and water resources, as well as conducting research and training. The data set published by \cite{Domenico_fao_data} contains a subset of the FAO international trade data for food products for the year 2010. The data can be described as a multiplex network on a set of countries as nodes, with layers in the multiplex corresponding to the individual food categories traded. In each layer, the directed edges are used to indicate the existence of a trade relationship, attributed with the weights that represent the quantities traded.

As the first step of data pre-processing, we remove the entries with unknown country of origin. Additionally, we merge data for four regions associated with China - mainland China, Taiwan, Hong Kong and China as a whole - since the records distinguish these entries. Considering records of Hong Kong, Taiwan and mainland China individually would be interesting from an application point of view as the three regions are economically and politically distinct, but due to the presence of records of China as a whole country, we chose to merge the records to avoid ambiguity since any trade with China could be recorded as either a transaction with the country of China, or with a specific region of it.

In this work, we focus on analysing the following four major food categories: dairy products (such as milk, cheese, butter, infant formula), fruit and vegetables (fresh, dried and prepared), grain products (flours, breads, whole grains, etc.) and meat and poultry (meats and edible offals, eggs, processed meats). These food categories correspond to the layers in the multiplex international food trade network analysed in Section \ref{Application}. Due to varying levels of participation by countries in the trade network; reflected in the number of trade partners and the values of imports and exports, we limit our analysis to the top 80 most active countries. These countries are defined as being among the top importers and exporters across the specified food categories.

We construct a multiplex network on the set of 80 chosen countries with four layers corresponding to the product categories. In each layer, we first assign directed edges connecting the pairs of exporting and importing countries in this particular category. We then attribute the weights to the existing edges departing from each country by computing the proportions of its exports to individual countries out of its total exports in the product category. Where no edge has been assigned, i.e. no trade in this product has occurred between a pair of countries, the corresponding weight is zero. In the resulting multiplex network, the sum of weights of outgoing edges for any country is one in each of the food categories, i.e. the proportions refer to the share of exports out of total exports for each country. In very rare cases where the country has not exported a particular food category, the sum of its export weights is zero for this product type. Note that no country can have a zero weight sum in all product categories as the set of countries has been chosen so that only the most active exporters and importers are included in the network. 

In an attempt to understand the similarities in food trade patterns and relative volumes traded between major food exporting countries, we perform clustering on the subset of FAO trade data described above using the methodology presented in Sections \ref{Dirichlet_SBM} and \ref{Extensions}. The main findings are outlined in Section \ref{Application}.

\section{Dirichlet stochastic block model} \label{Dirichlet_SBM}

Following \cite{dir_sbm}, this section briefly describes the main aspects of the Dirichlet stochastic block model (DirSBM).

Consider a fully-connected network on $n$ nodes with no self-loops and let $\mathbf{Y}$ be its weighted adjacency matrix, i.e. $y_{ij}>0$ for all $i \neq j$ and $y_{ii}=0$ for all $i$. Define $\mathbf{X}$ as the compositional adjacency matrix of the same network, i.e. $x_{ij}>0$ for $i\neq j$, $x_{ii}=0$ for all $i$ and $\sum_{l} x_{il}=1 ~\forall i$. The row entries of $\mathbf{X}$ are given by

\begin{equation}
    \mathbf{x}_i = \bigg( \frac{y_{i1}}{\sum_{j\neq i}^{n}y_{ij}},  \frac{y_{i2}}{\sum_{j\neq i}^{n}y_{ij}},\ldots,\frac{y_{i(i-1)}}{\sum_{j\neq i}^{n}y_{ij}},0,\frac{y_{i(i+1)}}{\sum_{j\neq i}^{n}y_{ij}},\ldots,\frac{y_{in}}{\sum_{j\neq i}^{n}y_{ij}} \bigg).
\end{equation}

The Dirichlet SBM models each compositional vector \\ $\mathbf{x}_i^*=(x_{i1},\ldots,x_{i(i-1)},x_{i(i+1)},\ldots,x_{in})$ as

\begin{equation} \label{x_distribution}
    \mathbf{x}_{i}^*\lvert\mathbf{z}_i, \mathbf{Z}_{-i} \sim Dir(\mathbf{z}_i \mathbf{A} {\mathbf{Z}}^{\top}_{-i}),
\end{equation}

\noindent where $\mathbf{A}$ is a $K\times K$ parameter matrix with strictly positive values, $K$ is the number of blocks or clusters, $\mathbf{z}_i$ is a binary cluster assignment vector for observation $i$, i.e. $z_{ik}=1$ when $i$ is a member of cluster $k$ and zero otherwise, and $\mathbf{Z}_{-i}$ is the $(N-1)\times K$ matrix formed by the binary cluster assignment vectors of the remaining $(n-1)$ nodes in the network.

To perform inference, a special variant of the classification expectation-maximisation algorithm \citep{Dempster-EM,ClassificationEM} from \cite{HybridML} is employed. We define a hybrid likelihood of the data based on a working independence assumption. That is, for each node $i$, we only consider the corresponding cluster label $\mathbf{z}_i$ as the true latent variable whilst assuming the fixed nature of the remaining partition, which we denote with $\widetilde{\mathbf{Z}}_{-i}$. This leads to a factorisation of the likelihood over the nodes, making inference possible. The complete data hybrid log-likelihood for DirSBM is given by

\begin{equation} \label{Complete_hybrid_ll}
\begin{aligned}
    l^{hyb}_c(\mathbf{A},\boldsymbol{\theta})& = \sum_{i=1}^{n} \sum_{k=1}^{K} z_{ik} \log p(\mathbf{x}_i|z_{ik}=1,\widetilde{\mathbf{Z}}_{-i}) + \sum_{i=1}^{n} \sum_{k=1}^{K} z_{ik} \log \theta_k\\& = 
    \sum_{i=1}^{n} \sum_{k=1}^{K} z_{ik} \Bigg [ \log \Gamma (\sum_{j\neq i}^{n} \Tilde{\alpha}_j) - \sum_{j\neq i}^{n} \log \Gamma (\Tilde{\alpha}_j) + \sum_{j\neq i}^{n} (\Tilde{\alpha_j}-1) \log x_{ij} \Bigg ] \\& \hspace{4mm} + \sum_{i=1}^{n} \sum_{k=1}^{K} z_{ik} \log \theta_k, 
\end{aligned}
\end{equation}

\noindent where $\Tilde{\alpha}_j=\sum_{h=1}^{K} \Tilde{z}_{jh} \alpha_{kh}$, with $\Tilde{z}_{jh}$ being the fixed cluster allocation of node $j$ in the network. The variant of the classification EM algorithm then proceeds by iterating between the following steps:

\vspace{5mm}
\noindent \textbf{Expectation (E) step:} At iteration $t$, node $i$ is assigned to cluster $k$ with probability

\begin{equation}
    \hat{z}_{ik}^{(t)} = \widehat{\Pr}(z_{ik}=1|\mathbf{x}_i,\widetilde{\mathbf{Z}}_{-i}^{(t-1)}) \propto \hat{\theta}_k^{(t-1)} \dfrac{\Gamma(\sum_{j\neq i}^{n} \Tilde{\alpha}_j^{(t-1)})}{\prod_{j\neq i}^{n} \Gamma (\Tilde{\alpha}_j^{(t-1)})} \prod_{j\neq i}^{n} x_{ij}^{\Tilde{\alpha}_j^{(t-1)}},
\end{equation}
\noindent with $\Tilde{\alpha}_j^{(t-1)}=\sum_{h=1}^{K} \Tilde{z}_{jh}^{(t-1)} \hat{\alpha}_{kh}^{(t-1)}$. 

\vspace{5mm}
\noindent\textbf{Classification (C) step:} The C-step is performed in a greedy fashion following \cite{HybridML}. The nodes are considered in a sequential manner and their cluster membership is decided based on the cluster assignment producing the highest value of the hybrid log-likelihood, i.e. for node $i$

\begin{equation} \label{c-step}
    c_i^{(t)} = \underset{k=1,\ldots,K}{\arg \max} \hspace{2mm} \sum_{l=1}^{n} \log \left( \sum_{h=1}^{K}  \hat{\theta}_h^{(t-1)} p(\mathbf{x}_l | c_l=k,\mathbf{c}_{-l}=\Tilde{\mathbf{c}}_{-l}) \right).
\end{equation}

\noindent $\mathbf{c}^{(t)}$ and $\widetilde{\mathbf{Z}}^{(t)}$ are updated after each node is assigned to a cluster, so any changes in the cluster assignment of node $i$ are propagated when the assignments for successive nodes are decided on.

\vspace{5mm}
\noindent\textbf{Maximisation (M) step:} The mixing proportions are updated using

\begin{equation} \label{m-step}
    \hat{\theta}_k^{(t)} = \frac{\sum_{i=1}^{n} \hat{z}_{ik}^{(t)}}{n}.
\end{equation}
The set of Dirichlet parameter estimates $\{\hat{\alpha}_{kh}^{(t)}\}_{k,h=1}^{K}$ are computed numerically using the R function $optim$ \citep{r_reference} using the L-BFGS-B optimisation routine \citep{Byrd_optim}. 

\section{DirSBM for multiplex networks} \label{Extensions}
In this section we outline the proposed extensions of DirSBM to clustering general multiplex networks with compositional edges.

\subsection{DirSBM for compositional networks with zeros} \label{Extensions_zeros}

The original DirSBM assumes that the network is fully connected, i.e. all edges are present between all pairs of nodes. In the case of networks with absent edges, to fit the DirSBM, any zero values are treated as zero-weighted edges. Subsequently, to address the issue of zero compositional parts, zeros in the raw count data are replaced with a small value, such as $0.001$ \citep{dir_sbm}. While this is a standard approach in compositional data analysis, it can potentially reduce the robustness of results, as these small values are factored into the estimation of the Dirichlet concentration parameters. Additionally, the network's density should be carefully considered when evaluating the suitability of this approach. In denser networks, the effect of adding small values to zero edges is minimized, leading to less impact on clustering outcomes and parameter estimation. However, this approach also assumes that the presence of zero compositional parts is independent of the clustering structure, a potentially restrictive and unrealistic assumption.

For these reasons, we propose an extension of the DirSBM to account for and model the presence or absence of edges, in the spirit of other stochastic block models \citep{multiplex_sbm_barbillon,weighted_sbm}, hence enabling clustering of sparse compositional networks. In this framework, each zero in the compositional vectors is treated as an indication of an absent edge between the corresponding pair of nodes in the network. 

The extended DirSBM, which models absent edges, has the following generative process:

\begin{enumerate}
    \item Let $\boldsymbol{\theta}=(\theta_1, \ldots, \theta_K)$ be a vector of cluster membership proportions, such that $\sum_{k=1}^{K} \theta_k = 1$. Assign each node $i$ to cluster $k$ with probability $\theta_k$, denoting by $\mathbf{c}$ the resulting cluster assignments vector.
    \item Let $\mathbf{P}=\{p_{kh}\}_{k,h=1}^{K}$ be a $K \times K$ matrix of connectivity probabilities. For each pair of nodes, generate binary edges according to
    $$
        e_{ij}|c_i=k,c_j=h \sim Bernoulli(p_{kh})
    $$
    and store them in the adjacency matrix denoted $\mathbf{E}$, with rows $\mathbf{e}_i$, $i=1,\ldots,n$.
    \item Let $\mathbf{A}=\{\alpha_{kh}\}_{k,h=1}^{K}$ be a $K \times K$ matrix with strictly positive values. Generate compositional observations as 
    $$
    \mathbf{x}_{i}^*\lvert \mathbf{e}_i, \mathbf{c} \sim Dir(\boldsymbol{\alpha}_i^*),
    $$
    where $\boldsymbol{\alpha}_i^*$ is defined as a vector of parameter values with entries $\alpha_{c_i c_j}$ only for existing edges, i.e. $\boldsymbol{\alpha}_i^*=(\alpha_{c_i c_j})$ where $i,j \in \{i,j \colon e_{ij}=1\}$.
\end{enumerate}

Note that in this generative procedure, $\mathbf{x}_{i}^{*}$ is the compositional weight vector that omits the entries corresponding to absent edges. Therefore such $\mathbf{x}_{i}^{*}$ have different dimensionalities, denoted as $d_i$, determined by the outgoing degree of node $i$. Let us denote by $\mathbf{x}_i$ the full $n$-dimensional vector of weights from node $i$ with zeros in places of absent edges, and by $\mathbf{X}$ an $(n \times n)$ matrix of edge weights (both present and absent) for the whole network.

Lastly, we note that this additional modelling step can also be used in combination with the zero treatment approach of the original DirSBM, where zeros are replaced with small values, if there is reason to believe that two distinct types of zeros exist in the data: one representing absent edges and the other representing present but zero-weighted edges. In such cases, the replacement strategy would be applied to zero-weighted edges during data preprocessing, ensuring that only truly absent edges are modeled as absent by the proposed model extension. For example, this combined approach could be useful in the analysis of international flights network. In such networks, zeros might indicate either the complete absence of flights between two airports or simply the lack of flights within a specific time window. Edges representing infrequent or seasonal flights could be treated as zero-weighted, distinguishing them from absent edges, where no flights would be observed regardless of the time window considered.

\subsubsection{Interpretation of model parameters} \label{Extensions_zeros_interpretation}

Similar to the original DirSBM, the Dirichlet concentration parameters are not measured on the same scale as the compositional data they describe, therefore, to aid interpretation we derive expressions for the expected node-to-node exchange shares and the expected cluster-to-cluster exchange shares. The former give us the proportions that we expect to observe between pairs of nodes given their cluster labels, and the latter aggregate the proportions that are exchanged between pairs of clusters collectively. These functions of Dirichlet concentration parameters are proportions themselves, so their interpretation is more intuitive than that of the raw values of the Dirichlet parameters.

The original model assumes that all connections between pairs of nodes are present, except for self-loops. This simplifies computations, as each node has exactly $(n_k-1)$ connections within its own cluster, where $n_k$ is the number of nodes in cluster $k$, $n_h$ connections with cluster $h$, where $n_h$ is the number of nodes in cluster $h$, and so on. This also means that the expected node-to-node shares are the same for nodes within the same cluster pair. However, when extending the model to account for the presence and absence of edges, this is no longer the case, as the number of connections differs for each node. We define $\mathbf{W}$ to be an $(n \times n)$ matrix of expected node-to-node exchange shares, with entry $w_{ij}$ corresponding to the expected proportion sent from node $i$ to node $j$:

\begin{equation} \label{W}
    w_{ij}=\frac{e_{ij}\alpha_{c_i c_j}}{\sum_{l=1}^{n}e_{il}\alpha_{c_i c_l}}.
\end{equation}

Note that Equation \eqref{W} refers to the expected proportions sent between any pair of nodes, therefore this is not $\mathbb{E}[\mathbf{x}_i^*\lvert \mathbf{e}_{i},\mathbf{c}]$, which only considers existing, non-zero edges - this also includes pairs that are not connected, i.e. where $e_{ij}=0$, and hence the expected proportions for such pairs are zero. Each row of matrix $\mathbf{W}$ sums to one.

As for the expected proportions exchanged between clusters, the expression again depends on the number of connections between nodes in the respective clusters. We can construct a $(K\times K)$ matrix $\mathbf{V}$, with entry $v_{kh}$ corresponding to the aggregated proportion sent by cluster $k$ to cluster $h$, out of the total for cluster $k$:

\begin{equation} \label{V}
    v_{kh}=\frac{\sum_{i=1}^{n}\sum_{j=1}^{n} e_{ij} \mathbbm{1}\{c_i=k\}\mathbbm{1}\{c_j=h\}\alpha_{c_i c_j}}{\sum_{i=1}^{n}\sum_{j=1}^{n} e_{ij} \mathbbm{1}\{c_i=k\} \alpha_{c_i c_j}}.
\end{equation}
Similarly to $\mathbf{W}$, each row of $\mathbf{V}$ sums to one. 

\subsubsection{Inference} \label{Extensions_zeros_inference}
To perform inference for DirSBM for compositional networks with zeros, we follow the same procedure as the original DirSBM based on hybrid maximum likelihood \citep{HybridML}, and derive a variant of the classification expectation-maximisation algorithm.

We begin by defining the hybrid log-likelihood as

\begin{equation} \label{obs_hybrid_ll}
 l^{hyb} (\mathbf{P},\mathbf{A},\boldsymbol{\theta}) = \sum_{i=1}^{n} \log \left ( 
 \sum_{k=1}^{K} \theta_k p(\mathbf{e}_i,\mathbf{x}_i \lvert c_i=k, \mathbf{c}_{-i} = \Tilde{\mathbf{c}}_{-i})
 \right ).
\end{equation}

Since the cluster labels are not observed, they are considered latent variables. Let $z_{ik}=1$ when node $i$ is a member of cluster $k$ and zero otherwise. Then the complete data hybrid log-likelihood of data $(\mathbf{E},\mathbf{X})$ and the set of latent variables $\mathbf{Z}$ is:

\begin{equation} \label{complete_hybrid_ll}
\begin{aligned}
    l_{c}^{hyb}(\mathbf{P},\mathbf{A},\boldsymbol{\theta})=\sum_{i=1}^{n} \sum_{k=1}^{K} z_{ik} \Bigg( &
    \log \theta_k + \sum_{j=1}^{n} \sum_{h=1}^{K} \Tilde{z}_{jh} (1-e_{ij}) \log (1-p_{kh}) + \sum_{j=1}^{n} \sum_{h=1}^{K} \Tilde{z}_{jh} e_{ij} \log p_{kh} \\ & + \log \Gamma (\sum_{j=1}^{n} e_{ij} \Tilde{\alpha}_{j}) - \sum_{j=1}^{n} e_{ij} \log \Gamma ( \Tilde{\alpha}_{j}) +\sum_{j=1}^{n} e_{ij} ( \Tilde{\alpha}_{j} - 1) \log x_{ij} \Bigg),
    \end{aligned}
\end{equation}

\noindent where $\Tilde{z}_{jh}=\mathbbm{1}\{\Tilde{c}_j=h\}$ is the fixed indicator of cluster allocation for node $j$ and $\Tilde{\alpha}_{j}=\sum_{h=1}^{K} \Tilde{z}_{jh} \alpha_{kh}$.

The steps of the classification EM algorithm are outlined below, and full details of derivations are available in Appendix \ref{app_CEM_steps}.

\vspace{5mm}
\noindent \textbf{E-step:} At iteration $t$, node $i$ is assigned to cluster $k$ with probability

\begin{equation}
\begin{aligned}
\hat{z}_{ik}^{(t)} & = \widehat{Pr}(z_{ik}=1 \lvert \mathbf{e}_i, \mathbf{x}_i,\widetilde{\mathbf{Z}}_{-i}^{(t-1)}) \\& \propto \hat{\theta}_k
\left[
\prod_{j=1}^{n}\prod_{h=1}^{K} \left[(1-\hat{p}_{kh})^{1-e_{ij}}\hat{p}_{kh}^{e_{ij}}\right]^{\Tilde{z}_{jh}}
\right]
\left[
\frac{\Gamma(\sum_{j=1}^{n} e_{ij} \Tilde{\alpha}_{j})}{\prod_{j=1}^{n}\Gamma(\Tilde{\alpha}_{j})^{e_{ij}}} \prod_{j=1}^{n} x_{ij}^{e_{ij} (\Tilde{\alpha}_{j}-1)}
\right],
\end{aligned}
\end{equation}

\noindent where the cluster label assignments as well as the parameter estimates come from the previous, $(t-1)$ iteration of the algorithm, i.e. $\Tilde{z}_{jh}=\Tilde{z}_{jh}^{(t-1)}$, $\hat{\theta}_k=\hat{\theta}_k^{(t-1)}$, $\hat{p}_{kh}=\hat{p}_{kh}^{(t-1)}$ and 
$\Tilde{\alpha}_{j}=\Tilde{\alpha}_{j}^{(t-1)}$.

\vspace{5mm}
\noindent \textbf{C-step:} Following closely the C-step of the original DirSBM based on \cite{HybridML}, a greedy approach is used. Every cluster label swap is considered for each node in a sequential manner and the nodes are assigned to the cluster with the highest hybrid log-likelihood value, i.e. 

\begin{equation} 
    c_i^{(t)} = \underset{k=1,\ldots,K}{\arg \max} \hspace{2mm} \sum_{l=1}^{n} \log \left( \sum_{h=1}^{K}  \hat{\theta}_h^{(t-1)} p(\mathbf{e}_l, \mathbf{x}_l | c_l=k,\mathbf{c}_{-l}=\Tilde{\mathbf{c}}_{-l}) \right),
\end{equation}

\noindent and then $\mathbf{c}^{(t)}$ and $\widetilde{\mathbf{Z}}^{(t)}$ are updated. Any changes in cluster labels are propagated through this step, i.e. if node $i$ is reassigned to a different cluster, such that $c_{i}^{(t)} \neq c_{i}^{(t-1)}$, we condition on its newly updated cluster label $c_{i}^{(t)}$ for any nodes that are considered after it.

\vspace{5mm}
\noindent \textbf{M-step:}
The updates for the mixing proportions are available in closed form and are given by
\begin{equation} 
    \hat{\theta}_k^{(t)} = \frac{\sum_{i=1}^{n} \hat{z}_{ik}^{(t)}}{n}.
\end{equation}

\noindent The connectivity matrix $\mathbf{P}$ is updated entry-wise using

\begin{equation} 
    \hat{p}_{kh}^{(t)} = \frac{\sum_{i=1}^{n} \sum_{j=1}^{n}\hat{z}_{ik}^{(t)}\Tilde{z}_{jh}^{(t)}e_{ij}}{\sum_{i=1}^{n} \sum_{j=1}^{n}\hat{z}_{ik}^{(t)}\Tilde{z}_{jh}^{(t)}},
\end{equation}
i.e. the number of edges that exist between cluster $k$ and cluster $h$, weighted by the probabilities of assigning the sender nodes to cluster $k$ divided by the total possible number of edges between cluster $k$ and cluster $h$ (weighted by probability of assigning the sender nodes to cluster $k$).

Following closely the estimation procedure of the original DirSBM, the estimate for the Dirichlet concentration parameters matrix $\mathbf{A}$ is found numerically using the R function \textit{optim} \citep{r_reference} using the L-BFGS-B optimisation routine \citep{Byrd_optim}.

The convergence criterion is analogous to that of the original DirSBM, hence the algorithm is considered to have reached convergence when 

\begin{equation}
    \left| \frac{l^{hyb}(\mathbf{\hat{P}}^{(t)},\mathbf{\hat{A}}^{(t)},\boldsymbol{\hat{\theta}}^{(t)})-l^{hyb}(\mathbf{\hat{P}}^{(t-1)},\mathbf{\hat{A}}^{(t-1)},\boldsymbol{\hat{\theta}}^{(t-1)})}{l^{hyb}(\mathbf{\hat{P}}^{(t)},\mathbf{\hat{A}}^{(t)},\boldsymbol{\hat{\theta}}^{(t)})} \right| < \epsilon.
\end{equation}
Throughout this work, we have used $\epsilon=10^{-4}$.

\subsection{Multiplex DirSBM (multi-DirSBM)} \label{Extensions_layers}

The key contribution of this work is the extension of the DirSBM to networks with multiple layers, each representing different types of interactions. Since multiplex networks are likely to exhibit a greater level of sparsity than single-layer networks, the extension to multiplex networks builds upon the DirSBM for compositional networks with zeros, as presented in Section \ref{Extensions_zeros}. We refer to the final proposed model, incorporating both extensions, as the multiplex Dirichlet stochastic block model (multi-DirSBM).

The data generating procedure for multi-DirSBM is a straightforward extension of the DirSBM for compositional networks with zeros presented at the start of Section \ref{Extensions_zeros}:

\begin{enumerate}
    \item Given a $K$-dimensional vector of mixing proportions $\boldsymbol{\theta}$, assign each node $i$ to cluster $k$ with probability $\theta_k$, and denote by $\mathbf{c}$ the resulting cluster assignments vector. Note that the partition into clusters is assumed to be constant across the layers of the network.

    \item Let $S$ be the number of layers in a multiplex network. Let $\mathcal{P}=\{\mathbf{P}^{(1)},\ldots,\mathbf{P}^{(s)}, \ldots, \mathbf{P}^{(S)}\}$ be a collection of $K\times K$ matrices of connectivity probabilities in each of the $S$ layers. Generate binary edges $e_{ij}^{(s)}$ between nodes $i$ and $j$ in layer $s$ using 
    $$
        e_{ij}^{(s)}|c_i=k,c_j=h \sim Bernoulli(p_{kh}^{(s)}).
    $$
    
    and store them in the set of adjacency matrices $\mathcal{E}=\{\mathbf{E}^{(1)},\ldots,\mathbf{E}^{(s)},\ldots,\mathbf{E}^{(S)}\}$.
    \item Given a collection of $K \times K$ parameter matrices with strictly positive values $\mathcal{A}=\{\mathbf{A}^{(1)},\ldots,\mathbf{A}^{(s)},\ldots,\mathbf{A}^{(S)}\}$, a vector of compositional weights for node $i$ in layer $s$ is generated from
    $$
    \mathbf{x}_{i}^{(s)*}\lvert \mathbf{e}_i^{(s)}, \mathbf{c} \sim Dir(\boldsymbol{\alpha}_i^{(s)*}),
    $$
    where $\boldsymbol{\alpha}_i^{(s)*}=(\alpha_{c_i c_j}^{(s)})$ with $i,j \in \{i,j \colon e_{ij}^{(s)}=1\}$.
\end{enumerate}

\subsubsection{Interpretation of model parameters} \label{Extensions_multiplex_interpretation}

The expressions of the expected node-to-node exchange shares and of the expected cluster-to-cluster exchange shares are similar to those presented in Section \ref{Extensions_zeros_interpretation}. To find the expected share sent by node $i$ to node $j$ in layer $s$, we compute

\begin{equation} \label{W_s}
    w_{ij}^{(s)}=\frac{e_{ij}^{(s)}\alpha_{c_i c_j}^{(s)}}{\sum_{l=1}^{n}e_{il}^{(s)}\alpha_{c_i c_l}^{(s)}}.
\end{equation}
In this way we can define a collection of $S$ $n\times n$ matrices of expected node-to-node exchange shares $\mathcal{W}=\{\mathbf{W}^{(1)},\ldots,\mathbf{W}^{(s)},\ldots,\mathbf{W}^{(S)}\}$.

For the expected proportions exchanged between clusters in layer $s$, we compute:

\begin{equation} \label{V_s}
    v_{kh}^{(s)}=\frac{\sum_{i=1}^{n}\sum_{j=1}^{n} e_{ij}^{(s)} \mathbbm{1}\{c_i=k\}\mathbbm{1}\{c_j=h\}\alpha_{c_i c_j}^{(s)}}{\sum_{i=1}^{n}\sum_{j=1}^{n} e_{ij}^{(s)} \mathbbm{1}\{c_i=k\} \alpha_{c_i c_j}^{(s)}},
\end{equation}
which can be used to construct a collection of $S$ $K\times K$ matrices of expected cluster-to-cluster exchange shares $\mathcal{V}=\{\mathbf{V}^{(1)},\ldots,\mathbf{V}^{(s)},\ldots,\mathbf{V}^{(S)}\}$. These quantities will be employed in Section~\ref{Application} to aid the interpretation of the results of multi-DirSBM applied to the FAO trade network data.

\subsubsection{Inference}
The inferential procedure for multi-DirSBM follows closely that one of DirSBM for compositional networks with zeros as outlined in Section \ref{Extensions_zeros_inference}, with one important modification. Although multi-DirSBM is a natural extension of the DirSBM in Section~\ref{Extensions_zeros_inference} to multiplex networks, care needs to be taken in specifying the likelihood for multi-DirSBM. DirSBM is designed for connected networks, therefore it cannot naturally handle isolated vertices or disconnected components. However, in the case of a multiplex network, to identify the clustering partition there is no need to require that each individual layer is a connected network, but only that the projection of the multiplex onto a single-layer network must be connected. Hence, the individual layers do not necessarily have to be connected on their own. Therefore, assuming independence between layers conditional on the latent variables, we propose a slight modification to the hybrid log-likelihood to account for cases where some nodes are disconnected or only receive incoming edges, thus being considered ``disconnected" in one direction (outgoing). The hybrid log-likelihood is as follows:

\begin{equation} \label{obs_hybrid_ll_multi}
    \begin{aligned}
        l^{hyb} (\mathcal{P},\mathcal{A},\boldsymbol{\theta}) & = \sum_{s=1}^{S}\sum_{i=1}^{n} \log \left ( 
        \sum_{k=1}^{K} \theta_k p(\mathbf{e}_i^{(s)}\lvert c_i=k, \mathbf{c}_{-i}=\Tilde{\mathbf{c}}_{-i})
        p(\mathbf{x}_i^{(s)}\lvert c_i=k, \mathbf{c}_{-i}=\Tilde{\mathbf{c}}_{-i},\mathbf{e}_i^{(s)})
        \right ) \\
        & = \sum_{s=1}^{S} \sum_{i=1}^{n} \log  
        \sum_{k=1}^{K} \theta_k 
        \left[
        \prod_{j=1}^{n}\prod_{h=1}^{K} \left[(1-p_{kh}^{(s)})^{1-e_{ij}^{(s)}}p_{kh}^{(s)e_{ij}^{(s)}}\right]^{\mathbb{I}\{c_i=k,\Tilde{c}_j=h\}}
        \right] \times \\
        & 
        \hspace{30mm}\left[
        \frac{\Gamma(\sum_{j=1}^{n}e_{ij}^{(s)}\Tilde{\alpha}_{j}^{(s)}+\mathbb{I}\{\sum_{j=1}^{n}e_{ij}^{(s)}=0\})}{\prod_{j=1}^{n}\Gamma(\Tilde{\alpha}_{j}^{(s)})^{e_{ij}^{(s)}}} \prod_{j=1}^{n} x_{ij}^{(s)e_{ij}^{(s)}(\Tilde{\alpha}_{j}^{(s)}-1)}
        \right]^{\mathbb{I}\{c_i=k\}},
    \end{aligned}
\end{equation}
where $\Tilde{\alpha}_j^{(s)}=\sum_{h=1}^{K} \Tilde{z}_{jh} \alpha_{kh}^{(s)}.$ In comparison to the single network counterpart of Section\ref{Extensions_zeros}, we introduce the additional term $\mathbb{I}\{\sum_{j=1}^{n}e_{ij}=0\}$ in the normalising constant of the Dirichlet probability density in Equation \eqref{obs_hybrid_ll_multi}. Whenever a node is isolated or only acts as a receiver in a specific layer, the likelihood contribution from such observations comes solely from the Bernoulli trials associated with present edges. By using $\mathbb{I}\{\sum_{j=1}^{n}e_{ij}=0\}$ as an additional term in the normalising constant of the Dirichlet probability density, we can ``switch off'' the contribution of the compositional edge weights vectors for nodes that do not act as senders in certain layers. This is because

\begin{equation*}
    p(\mathbf{x}_i^{(s)}\lvert c_i=k, \mathbf{c}_{-i}=\Tilde{\mathbf{c}}_{-i},\mathbf{e}_i^{(s)}) =
    \begin{cases}
        \frac{\Gamma(\sum_{j=1}^{n}e_{ij}^{(s)}\Tilde{\alpha}_{j}^{(s)})}{\prod_{j=1}^{n}\Gamma(\Tilde{\alpha}_{j}^{(s)})^{e_{ij}^{(s)}}} \prod_{j=1}^{n} x_{ij}^{(s)e_{ij}^{(s)}(\Tilde{\alpha}_{j}^{(s)}-1)}, \text{ when some $e_{ij}^{(s)}=1$} \vspace{3mm} \\
        1, \text{ when $e_{ij}^{(s)}=0 ~~ \forall j$}.
    \end{cases}
\end{equation*}

The introduction of latent cluster allocation variables leads to the complete data hybrid log-likelihood:

\begin{equation} \label{complete_hybrid_ll_multi}
\begin{aligned}
    l_{c}^{hyb} (\mathcal{P},\mathcal{A},\boldsymbol{\theta})= \sum_{s=1}^{S} \sum_{i=1}^{n} \sum_{k=1}^{K} z_{ik} \Bigg( &
    \log \theta_k + \sum_{j=1}^{n} \sum_{h=1}^{K} \Tilde{z}_{jh} (1-e_{ij}^{(s)}) \log (1-p_{kh}^{(s)}) + \sum_{j=1}^{n} \sum_{h=1}^{K} \Tilde{z}_{jh} e_{ij}^{(s)} \log p_{kh}^{(s)} \\ & + \log \Gamma \left(\sum_{j=1}^{n} e_{ij}^{(s)} \Tilde{\alpha}_{j}^{(s)} +\mathbb{I}\{\sum_{j=1}^{n}e_{ij}^{(s)}=0\}\right) - \sum_{j=1}^{n} e_{ij}^{(s)} \log \Gamma (\Tilde{\alpha}_{j}^{(s)}) \\& +\sum_{j=1}^{n} e_{ij}^{(s)} (\Tilde{\alpha}_{j}^{(s)} - 1) \log x_{ij}^{(s)} \Bigg).
    \end{aligned}
\end{equation}

The steps of the classification EM algorithm are provided below; Appendix \ref{app_CEM_steps} contains the full derivations. The algorithm is initialised from a number of random cluster partitions, subsequently, the run leading to the highest hybrid log-likelihood value at convergence is retained. This simple approach has been found to work well in practice in \cite{dir_sbm} in this context. In the description of the algorithm below, the superscript $(t)$ denoting the iteration counter has been omitted from the expressions to simplify the notation.

\vspace{5mm}
\noindent \textbf{E-step:} At iteration $t$, node $i$ is assigned to cluster $k$ with probability 

\begin{equation}
\begin{aligned}
\hat{z}_{ik} & = \widehat{Pr}(z_{ik}=1 \lvert \mathbf{e}_i^{(s)}, \mathbf{x}_i^{(s)},\widetilde{\mathbf{Z}}_{-i}) \\& \propto \hat{\theta}_k
\left[
\prod_{s=1}^{S}\prod_{j=1}^{n}\prod_{h=1}^{K} \left[(1-\hat{p}_{kh}^{(s)})^{1-e_{ij}}\hat{p}_{kh}^{(s)e_{ij}}\right]^{\Tilde{z}_{jh}}
\right]
\left[
\frac{\Gamma(\sum_{j=1}^{n} e_{ij} \Tilde{\alpha}_{j}^{(s)})}{\prod_{j=1}^{n}\Gamma(\Tilde{\alpha}_{j}^{(s)})^{e_{ij}}} \prod_{j=1}^{n} x_{ij}^{(s)e_{ij}^{(s)} (\Tilde{\alpha}_{j}^{(s)}-1)}
\right],
\end{aligned}
\end{equation}

\noindent calculated using the cluster assignments and the parameter estimates from the $(t-1)$ iteration of the algorithm.

\vspace{5mm}
\noindent \textbf{C-step:} Considering all possible cluster label swaps, node $i$ is assigned to the cluster with the highest hybrid log-likelihood value, i.e. 

\begin{equation} \label{c-step_multi}
    c_i = \underset{k=1,\ldots,K}{\arg \max} \hspace{2mm} \sum_{s=1}^{S} \sum_{l=1}^{n} \log \left( \sum_{h=1}^{K}  \hat{\theta}_h p(\mathbf{e}_l^{(s)}, \mathbf{x}_l^{(s)} | c_l=k,\mathbf{c}_{-l}=\Tilde{\mathbf{c}}_{-l}) \right),
\end{equation}

\noindent using the parameter estimates from the $(t-1)$ iteration of the algorithm; the corresponding entries of $\mathbf{c}$ and $\widetilde{\mathbf{Z}}$ are updated accordingly. The process is implemented for all nodes sequentially, conditional on the cluster labels of any preceding node.

\vspace{5mm}
\noindent \textbf{M-step:} At iteration $t$, the mixing proportions are updated as
\begin{equation} \label{theta-m-step_multi}
    \hat{\theta}_k = \frac{\sum_{i=1}^{n} \hat{z}_{ik}}{n},
\end{equation}

\noindent where the estimated probabilities $\hat{z}_{ik}$ come from the current iteration $t$.

For each of the $S$ layers, the connectivity matrices $\mathbf{P}^{(s)}$ are updated entry-wise using

\begin{equation} \label{P-m-step_multi}
    \hat{p}_{kh}^{(s)} = \frac{\sum_{i=1}^{n} \sum_{j=1}^{n}\hat{z}_{ik}\Tilde{z}_{jh}e_{ij}^{(s)}}{\sum_{i=1}^{n} \sum_{j=1}^{n}\hat{z}_{ik}\Tilde{z}_{jh}},
\end{equation}

\noindent where both the cluster assignment probabilities, $\hat{z}_{ik}$, and the classification labels, $\Tilde{z}_{jh}$, are from the current iteration $t$.

For each layer individually, the Dirichlet concentration matrix $\mathbf{A}^{(s)}$ is estimated numerically using the R function \textit{optim} \citep{r_reference} using L-BFGS-B optimisation routine \citep{Byrd_optim}.

The algorithm iterates the steps above until the following convergence criterion is met:

\begin{equation}
    \left| \frac{l^{hyb}(\hat{\mathcal{P}}^{(t)},\hat{\mathcal{A}}^{(t)},\hat{\boldsymbol{\theta}}^{(t)})-l^{hyb}(\hat{\mathcal{P}}^{(t-1)},\hat{\mathcal{A}}^{(t-1)},\hat{\boldsymbol{\theta}}^{(t-1)})}{l^{hyb}(\hat{\mathcal{P}}^{(t)},\hat{\mathcal{A}}^{(t)},\hat{\boldsymbol{\theta}}^{(t)})} \right| < \epsilon.
\end{equation}
The tolerance value used in this work is $\epsilon=10^{-4}$.

\subsection{Model selection} \label{model_selection}

The model extensions proposed above require the number of clusters $K$ to be specified in advance. With the true number of clusters usually being unknown, a model selection criterion is needed in order to choose the underlying number of clusters appropriately. Following the original DirSBM, the integrated completed likelihood criterion \citep[ICL,][]{ICL,Daudin_mm_random_graphs} has been tested for the purpose of model selection. ICL is a popular model selection criterion in stochastic block modelling that has been used to choose the number of clusters in works such as \cite{multiplex_sbm_barbillon}, \cite{sbm_interaction_lengths} and \cite{weighted_sbm}. The ICL for multi-DirSBM is given by

\begin{equation}
    ICL(K) = l^{hyb}_c(\hat{\mathcal{P}},\hat{\mathcal{A}},\boldsymbol{\hat{\theta}}|\mathcal{X},\mathcal{E},\mathbf{\hat{Z}}) - K^2 S \log [S n(n-1)] -\frac{1}{2}(K-1)\log n,
\end{equation}

\noindent where $l^{hyb}_c(\hat{\mathcal{P}},\hat{\mathcal{A}},\boldsymbol{\hat{\theta}}|\mathcal{X},\mathcal{E},\mathbf{\hat{Z}})$ is the complete data hybrid log-likelihood of Equation \eqref{complete_hybrid_ll_multi}, used in place of the complete data log-likelihood as described in \cite{HybridML} and evaluated at the convergence of the algorithm. Appendix \ref{app_icl} provides details on the derivation of the criterion above that follow closely those in \cite{Daudin_mm_random_graphs}.

However, the simulation study in Section \ref{model_selection_performance} revealed that the use of the ICL for model selection produces mixed results, with the tendency to underestimate the number of clusters, especially in cases involving data with a larger number of clusters. 
To address this issue and inspired by the discussion on model selection in \cite{HybridML}, a Bayesian information criterion \citep[BIC,][]{bic} is considered for model selection, resulting in improved performance, as demonstrated in Section \ref{model_selection_performance}. The standard BIC cannot be directly employed to select the number of clusters in SBMs because the marginal likelihood is intractable. To overcome this challenge, we leverage the hybrid log-likelihood employed in multi-DirSBM estimation. Following \cite{HybridML}, we define a criterion based on the optimized hybrid log-likelihood. Specifically, the BIC for multi-DirSBM is given by:

\begin{equation}
    BIC(K) = l^{hyb}(\hat{\mathcal{P}},\hat{\mathcal{A}},\boldsymbol{\hat{\theta}}|\mathcal{X},\mathcal{E},\mathbf{\hat{Z}}) - K^2 S \log [S n(n-1)],
\end{equation}

\noindent where $l^{hyb}(\hat{\mathcal{P}},\hat{\mathcal{A}},\boldsymbol{\hat{\theta}}|\mathcal{X},\mathcal{E},\mathbf{\hat{Z}})$ is the hybrid log-likelihood from Equation \eqref{obs_hybrid_ll_multi} evaluated at the convergence of the algorithm in the spirit of \cite{HybridML}. The results of using both ICL and BIC for model selection are compared in Section \ref{model_selection_performance}. As the BIC has shown superior performance in selecting the correct number of clusters in the conducted simulation study, we conclude that it is a more suitable model selection criterion for multi-DirSBM.

Lastly, we note that both the above criteria can be used to perform model selection for DirSBM for compositional networks with zeros as presented in Section \ref{Extensions_zeros} by simply setting $S=1$.

\section{Simulation studies} \label{Sim_studies}

In this section, we evaluate the performance of multi-DirSBM on synthetic data by examining its clustering accuracy, parameter estimation quality, and model selection capabilities across various scenarios. These results are compared against benchmark stochastic block models (SBMs).

The synthetic data are generated on the basis of the sampling model described at the beginning of Section \ref{Extensions_layers}. As described in \cite{dir_sbm}, this data generating process can be equivalently formulated as generating a collection of independent Gamma-distributed samples, which are later normalised and converted into compositions. By extending this approach to multi-DirSBM, the compositional multidimensional network data can be generated according to the procedure described in Appendix \ref{app_data_gen_gamma}. We implement this approach to generate the synthetic networks, to have access to both the raw weighted and the composition-weighted data. The raw weighted multiplex data will be used as described in Section~\ref{sec:clust_perf}.

The difficulty in accurately recovering the latent structure, estimating parameters, and determining the number of clusters depends on several factors:

\begin{itemize}
    \item Number of nodes, $n$;
    \item Number of clusters, $K$;
    \item Number of layers in the multiplex network, $S$;
    \item Density of the network, i.e. how large the connectivity probabilities in $\mathbf{P}^{(s)}$ are;
    \item Cluster overlap, i.e. how similar the connectivity probabilities $p_{kh}^{(s)}$ in the individual layers are;
    \item Level of Dirichlet concentration parameter homogeneity, i.e. how similar the entries in the rows of $\mathbf{A}^{(s)}$ are.
\end{itemize}

The effect of Dirichlet concentration parameter homogeneity on the performance of DirSBM has been extensively studied in \cite{dir_sbm} for networks of varying sizes and cluster counts. Consequently, in this work, we set the Dirichlet parameter homogeneity to a medium level across all scenarios and focus on factors intrinsic to multi-DirSBM, namely network density and cluster overlap. To genereae synthetic data, we consider the combinations given in Table \ref{tab_settings}.

\setlength{\extrarowheight}{2pt}
\setlength{\tabcolsep}{3pt}
\begin{table}[t]
    \caption{Scenario settings for synthetic data generation to evaluate multi-DirSBM performance.}
    \label{tab_settings}
    \centering
    \begin{tabular}{c|c|c|c|c}
   $n$ & $K$ &$S$ & Network density & Cluster overlap \\
    \hline
    50 & 2 & 2 & High & No \\
    50 & 2 & 4 & Low & Yes \\
    50 & 3 & 2 & High & Yes \\
    50 & 3 & 4 & Low & No \\
    100 & 3 & 2 & High & No \\
    100 & 3 & 4 & High & Yes \\
    100 & 5 & 2 & High & No \\
    100 & 5 & 4 & Low & Yes \\
    \end{tabular}
\end{table}

The parameter matrices used to generate the data are given in Appendix \ref{app_sim_params}. For each set of combinations above, we generate 20 synthetic multiplex networks. When fitting multi-DirSBM, to initialise the algorithm, we employ five initial random cluster partitions and select the best run yielding the highest value of the hybrid log-likelihood.

\subsection{Clustering performance}\label{sec:clust_perf}

In order to assess the clustering performance of multi-DirSBM, we propose comparing it to benchmark clustering methods that are often used in practice. 

The first method we compare multi-DirSBM to is k-means \citep{kmeans} on raw continuos data. To construct the data matrix for k-means, we take the raw weighted adjacency matrices for each of the layers (i.e. the ones used to compute the compositions rather than the compositional data themselves), each of dimensionality $n \times n$, and stack them together to form a large $Sn \times n$ data matrix. Note that this is only possible if the raw continuous data are available, which may not be the case in some applications. This approach allows us to examine the importance of analysing the data accounting for the dependencies between the nodes, as well as analysing the data in their compositional format as opposed to the raw data.

The second competitor employed in this study is a binary SBM \citep{Nowichi2001}. In many applications, it is common practice to aggregate information from multiple types of interactions between nodes to construct a single-layer network. For instance, in a social network, various relationships such as family ties, friendships, and co-working connections can be combined into a single network indicating whether two individuals are acquainted. Similarly, in our case, we construct a one-layer binary network by assigning an edge between two nodes if an edge exists in at least one layer of the multiplex network. Once such a network is obtained, we fit a standard Bernoulli SBM using the $blockmodels$ package in R \citep{blockmodels_package}.

The final model we consider is the multiplex SBM by \cite{multiplex_sbm_barbillon}. From the original weighted multiplex network, we derive a binary multiplex network by constructing a collection of binary adjacency matrices that indicate the presence or absence of edges between nodes in each layer. We then fit the binary multiplex stochastic block model to this binary representation of the multiplex using the $blockmodels$ package \citep{blockmodels_package}. The comparison between the multiplex SBM and the multi-DirSBM allows the investigation of the impact of incorporating edge weight information when performing clustering.

\setlength{\extrarowheight}{2pt}
\setlength{\tabcolsep}{4pt}
\begin{table}[!t]
    \caption{Comparison of clustering performance of k-means on original continuous data, binary SBM, multiplex SBM and multi-DirSBM, evaluated on synthetic data with varying settings: number of nodes $n$, numbers of clusters $K$, numbers of layers $S$, network densities (high or low) and cluster overlap (yes and no). The results are presented as mean adjusted Rand index values, with standard deviation in parentheses, computed across 20 synthetic data sets.}
    \label{tab_ari}
    \centering
    \begin{tabular}{ccccc||cccc}
    \multicolumn{5}{c||}{Scenario settings} & \multicolumn{3}{c}{Model} \\
    \hline \hline
    $n$ & $K$ & $S$ & Density & Overlap & k-means & SBM & Multiplex SBM & multi-DirSBM \\
    \hline
    50 & 2 & 2 & High & No & 0.949 (0.227) & 1 (0) & 1 (0) & 1 (0) \\
    50 & 2 & 4 & Low & Yes & 0.937 (0.226) & 0.384 (0.358) & 0.905 (0.295) & 1 (0) \\
    50 & 3 & 2 & High & Yes & 0.629 (0.201) & 0.627 (0.217) & 0.979 (0.0408) & 1 (0) \\
    50 & 3 & 4 & Low & No & 0.671 (0.207) & 0.0739 (0.130) & 0.922 (0.191) & 1 (0) \\
    100 & 3 & 2 & High & No & 0.785 (0.248) & 1 (0) & 1 (0) & 1 (0) \\
    100 & 3 & 4 & High & Yes & 0.789 (0.265) & 1 (0) & 1 (0) & 1 (0) \\
    100 & 5 & 2 & High & No & 0.755 (0.151) & 1 (0) & 1 (0) & 0.993 (0.0323) \\
    100 & 5 & 4 & Low & Yes & 0.696 (0.198) & 0.0510 (0.0745) & 0.977 (0.0697) & 0.987 (0.0562) \\
    \end{tabular}
\end{table}

We note that all the models are implemented fixing the number of clusters to the actual value used to generate the data. Table \ref{tab_ari} presents the mean adjusted Rand index \citep[ARI,][]{adjRand} computed on 20 synthetic networks (standard deviations in brackets) of the different clustering methods for the scenarios in Table \ref{tab_settings}. The ARI measures the level of agreement between two partitions, accounting for agreement by chance. It takes a value of one when the two partitions are in perfect agreement, while values close to zero denote disagreement.

As observed in Table \ref{tab_ari}, k-means attains overall inferior performance compared to the other methods that account for the node-dependence in the data. It sometimes outperforms the binary SBM, but its performance tends to be rather volatile as indicated by consistently large standard deviations. The binary SBM (applied on the single network derived from the multiplex) tends to perform well when the the network layers have high density, but it performs poorly when the connections in the layers are sparse. This could be due to the fact that, in dense multiplex networks, most pairs of nodes are connected across multiple layers. Hence, a single network constructed from such a multiplex will itself be dense, with absent edges being more informative for clustering than present edges. This is because the occurrence of absent edges will be rare, since for an edge to be absent in the aggregated network it would need to be absent across all layers of the multiplex. On the other hand, in a sparse multiplex, a single-layer projection can still be dense since an edge is assigned if it exists in any layer. This may result in masking the clustering structures in the layers. The multiplex SBM successfully recovers the latent structure across most scenarios. However, its performance declines when the number of layers is higher, the network density is low, and there the clusters overlap. The proposed multi-DirSBM achieves near perfect clustering performance, with a mean ARI of $1$ in all but two scenarios, , where only a single instance in each shows slightly reduced performance. Multi-DirSBM consistently achieves satisfactory clustering performance and outperforms the other methods across the scenarios considered.

\subsection{Parameter estimation quality} \label{param_estimation}

In this section, we assess the quality of parameter estimation of multi-DirSBM in the scenarios introduced at the start of Section \ref{Sim_studies}. Note that we only consider models with the number of clusters equal to the true $K$. For the assessment, for each synthetic multiplex, we compute the Frobenius distance between actual and estimated probability connectivity matrices and Dirichlet concentration matrix parameters for each layer. These distances are then averaged across layers to provide a general measure of quality of the parameter estimates for each multiplex. Table \ref{tab_distance} presents the mean layer Frobenius distance \citep{matrix_computations} averaged across the 20 synthetic multiplex networks (standard deviations in brackets). 

\setlength{\extrarowheight}{4pt}
\setlength{\tabcolsep}{7pt}
\begin{table}[!t]
    \caption{Frobenius distance between true parameters, $\mathcal{P}$ and $\mathcal{A}$, and their estimates, $\hat{\mathcal{P}}$ and $\hat{\mathcal{A}}$, respectively, calculated from synthetic data with different numbers of nodes $n$, numbers of clusters $K$, numbers of layers $S$, network densities (high or low) and cluster overlap (yes and no). The  entries are the mean Frobenius distances over $S$ layers, with standard deviation in brackets, averaged across 20 synthetic data sets.}
    \label{tab_distance}
    \centering
    \begin{tabular}{ccccc||cc}
    \multicolumn{5}{c||}{Scenario settings} & \multicolumn{2}{c}{Mean Frobenius distance (standard deviation)} \\
    \hline \hline
    n & K & S & Density & Overlap & $S^{-1}\sum_{s=1}^{S} || \mathbf{P}^{(s)} - \hat{\mathbf{P}}^{(s)} ||_F$ & $S^{-1}\sum_{s=1}^{S} || \mathbf{A}^{(s)} - \hat{\mathbf{A}}^{(s)} ||_F$ \\
    \hline
    50 & 2 & 2 & High & No & 0.0387 (0.0149) & 0.0926 (0.0448)\\
    50 & 2 & 4 & Low & Yes & 0.0403 (0.0163) & 0.118 (0.0596)\\
    50 & 3 & 2 & High & Yes & 0.0802 (0.0222) & 0.185 (0.0504) \\
    50 & 3 & 4 & Low & No & 0.0816 (0.0226) & 0.298 (0.112)\\
    100 & 3 & 2 & High & No & 0.0402 (0.00964) & 0.111 (0.0323)\\
    100 & 3 & 4 & High & Yes & 0.0420 (0.0105) & 0.0926 (0.0247)\\
    100 & 5 & 2 & High & No & 0.150 (0.149) & 0.403 (0.247)\\
    100 & 5 & 4 & Low & Yes & 0.170 (0.169) & 0.484 (0.301)\\
    \end{tabular}
\end{table}

Table \ref{tab_distance} shows that, as the number of nodes increases, the parameter estimates of both connectivity probability and Dirichlet concentration parameters become more accurate. This is because a larger number of data points are available to estimate each individual parameter. Additionally, when the network density increases, the parameter estimates for $\mathcal{A}$ improve for the same reason. The largest mean Frobenius distance and standard deviations are observed for $K=5$, which is to be expected, as also in \cite{dir_sbm} it has been observed that parameter estimation becomes more challenging as the number of clusters increases. In general, the algorithm implemented to fit multi-DirSBM provides parameter estimates of satisfactory quality, and variations in the quality of the estimates in multiplex networks are consistent with those observed in single compositional networks, as reported in \cite{dir_sbm}.

\subsection{Model selection performance} \label{model_selection_performance}

To assess the model selection performance of the ICL and the BIC derived in Section \ref{model_selection}, we fit multi-DirSBM with different numbers of clusters to synthetic data generated under various regimes, as specified in Table \ref{tab_settings}, and compute the corresponding ICL and BIC values. When the true number of clusters $K$ is 2 or 3, we fit the models with up to 4 clusters, and when the actual number of clusters in the data is 5, we fit models with up to 6 clusters. Table \ref{tab_icl} presents the results of the simulation study evaluating the effectiveness of ICL for model selection, while Table \ref{tab_bic} shows the number of times each model was selected by the BIC.

\setlength{\extrarowheight}{2pt}
\setlength{\tabcolsep}{10pt}
\begin{table}[!t]
    \caption{Choice of number of clusters $K$ based on the integrated completed likelihood (ICL) value, for synthetic data with different numbers of nodes $n$, numbers of layers $S$, network densities (high or low), cluster overlap (yes and no) and numbers of clusters $K$. The entries indicate the number of times each number of clusters, $\hat{K}$, was selected by the criterion out of 20 simulations for each combination. Entries in \textbf{bold} indicate the number of times the correct number of clusters was selected, and \underline{underlined} entries indicate the most frequently selected number of clusters.}
    \label{tab_icl}
    \centering
    \begin{tabular}{ccccc||cccccc}
    \multicolumn{5}{c||}{Scenario settings} & \multicolumn{6}{c}{$\hat{K}$} \\
    \hline \hline
    n & S & Density & Overlap & K & 1 & 2 & 3 & 4 & 5 & 6 \\
    \hline
    50 & 2 & High & No & 2 & 4 & \underline{\textbf{13}} & 3 & & &\\
    50 & 4 & Low & Yes & 2 & 5 & \underline{\textbf{12}} & 3 & & & \\
    50  & 2 & High & Yes & 3 & 4 & 7 & \underline{\textbf{8}} & 1 & & \\
    50 & 4 & Low & No & 3 & 1 & 4 & \underline{\textbf{15}} & & & \\
    100 & 2 & High & No & 3 & & 4 & \underline{\textbf{11}} & 5 & & \\
    100 & 4 & High & Yes & 3 & 2 & 6 & \underline{\textbf{8}} & 4 & & \\
    100 & 2 & High & No & 5 & 5 & \underline{7} & 3 & 2 & \textbf{2} & 1 \\
    100 & 4 & Low & Yes & 5 & 2 & 2 & \underline{9} & 5 & \textbf{1} & 1  \\
    \end{tabular}
\end{table}

\setlength{\extrarowheight}{2pt}
\setlength{\tabcolsep}{10pt}
\begin{table}[!t]
    \caption{Choice of number of clusters $K$ based on the Bayesian information criterion (BIC) value, for synthetic data with different numbers of nodes $n$, numbers of layers $S$, network densities (high or low), cluster overlap (yes and no) and numbers of clusters $K$. The entries indicate the number of times each number of clusters, $\hat{K}$, was selected by the criterion out of 20 simulations for each combination. Entries in \textbf{bold} indicate the number of times the correct number of clusters was selected, and \underline{underlined} entries indicate the most frequently selected number of clusters.}
    \label{tab_bic}
    \centering
    \begin{tabular}{ccccc||cccccc}
    \multicolumn{5}{c||}{Scenario settings} & \multicolumn{6}{c}{$\hat{K}$} \\
    \hline \hline
    n & S & Density & Overlap & K & 1 & 2 & 3 & 4 & 5 & 6 \\
    \hline
    50 & 2 & High & No & 2 &  & \underline{\textbf{20}} &  & & &\\
    50 & 4 & Low & Yes & 2 &  & \underline{\textbf{20}} &  & & & \\
    50  & 2 & High & Yes & 3 &  &  & \underline{\textbf{20}} &  & & \\
    50 & 4 & Low & No & 3 &  &  & \underline{\textbf{20}} & & & \\
    100 & 2 & High & No & 3 & &  & \underline{\textbf{20}} &  & & \\
    100 & 4 & High & Yes & 3 &  &  & \underline{\textbf{20}} &  & & \\
    100 & 2 & High & No & 5 &  &  &  &  & \underline{\textbf{19}} & 1 \\
    100 & 4 & Low & Yes & 5 & & &  & 3 & \underline{\textbf{17}} &   \\
    \end{tabular}
\end{table}

According to Table \ref{tab_icl}, the ICL tends to select the correct number of clusters when the number of layers is high and the number of clusters is relatively low. This behaviour alings with the model selection performance of ICL in the original DirSBM \citep{dir_sbm}, making it an expected result. Performance of ICL on sparse networks appears to be better than on dense networks. Overall, performance of ICL in recovering the true number of clusters is mixed, especially in scenarios with high network density, a large number of clusters, a low number of layers. In such cases, ICL has a tendency to underestimate $K$.

In contrast, Table \ref{tab_bic} shows that the BIC selects the correct number of clusters in the vast majority of cases. However, in scenarios with with a larger number of clusters, it occasionally favours models with either too few or too many clusters. Despite this, the BIC consistently outperforms ICL in selecting the correct number of clusters. For this reason, we use the BIC to choose the number of clusters in the analysis of the international food exports data in Section \ref{Application}.

\section{Application of multi-DirSBM to FAO trade data} \label{Application}

This section discusses the main findings of fitting multi-DirSBM to the FAO trade data introduced in Section \ref{FAO_data}. We fitted models with $K=1,\ldots,7$, and the BIC selected a 5-cluster model as the best one. We note that the 5-cluster model provided a BIC value of 35,027, while the second best model with 6 clusters had a BIC value of 34,936.

\begin{figure}[htp] 
\centering
\begin{subfigure}{1\textwidth}
\includegraphics[width=1\linewidth,valign=t]{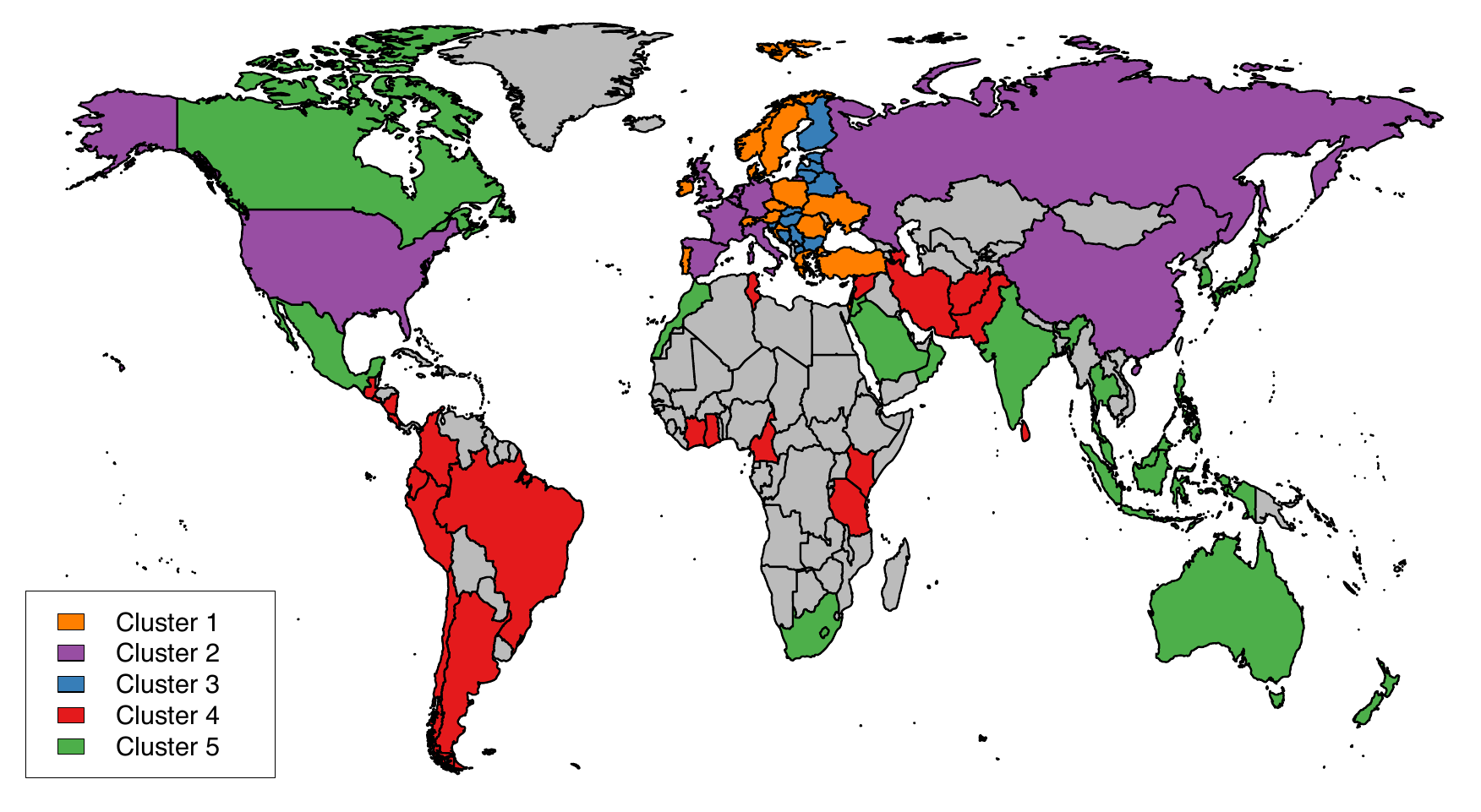}
\subcaption{}
\label{fao_data_map_K=5}
\end{subfigure}
\begin{subfigure}{1\textwidth}
\centering
\includegraphics[width=0.87\linewidth,valign=t]{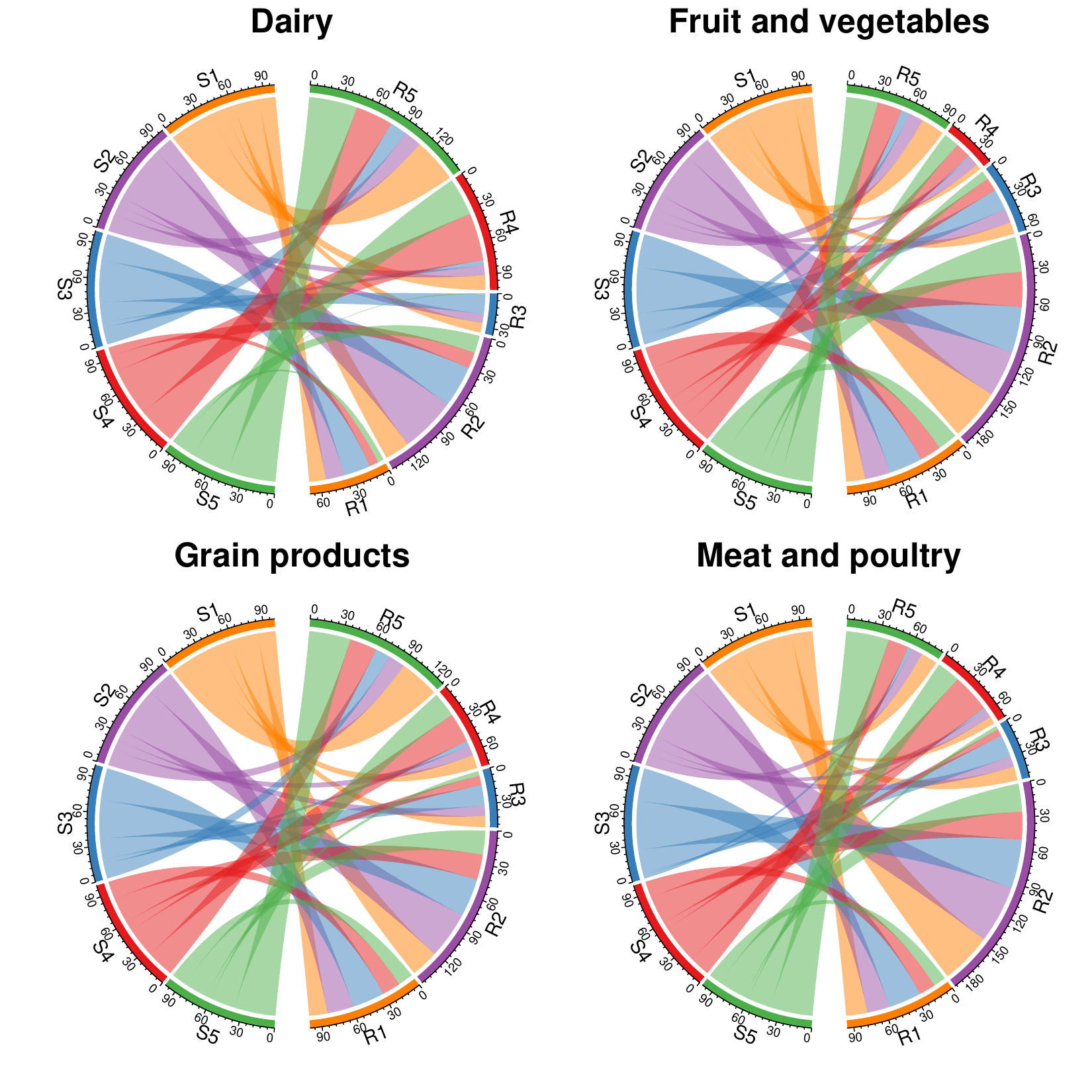}
\subcaption{}
\label{fao_data_cluster_shares_K=5}
\end{subfigure}
\caption{Clustering solution of multi-DirSBM with 5 clusters: \subref{fao_data_map_K=5} shows a world map with countries color-coded by cluster assignment. Countries shown in grey were not included in the analysis. \subref{fao_data_cluster_shares_K=5} chord diagram of total percentage flows between clusters of countries based on the expected cluster-to-cluster exchange shares matrix $\mathbf{\hat{V}}$. On the left, S1-S5 denote the sending clusters (1-5). Percentages departing from the sender sum up to 100\%. On the right, R1-R5 denote the receiving clusters (1-5).}
\label{Erasmus_K=5}
\end{figure}

\begin{figure}[p] 
\begin{subfigure}{1\textwidth}
\centering
\includegraphics[width=0.75\linewidth,valign=t]{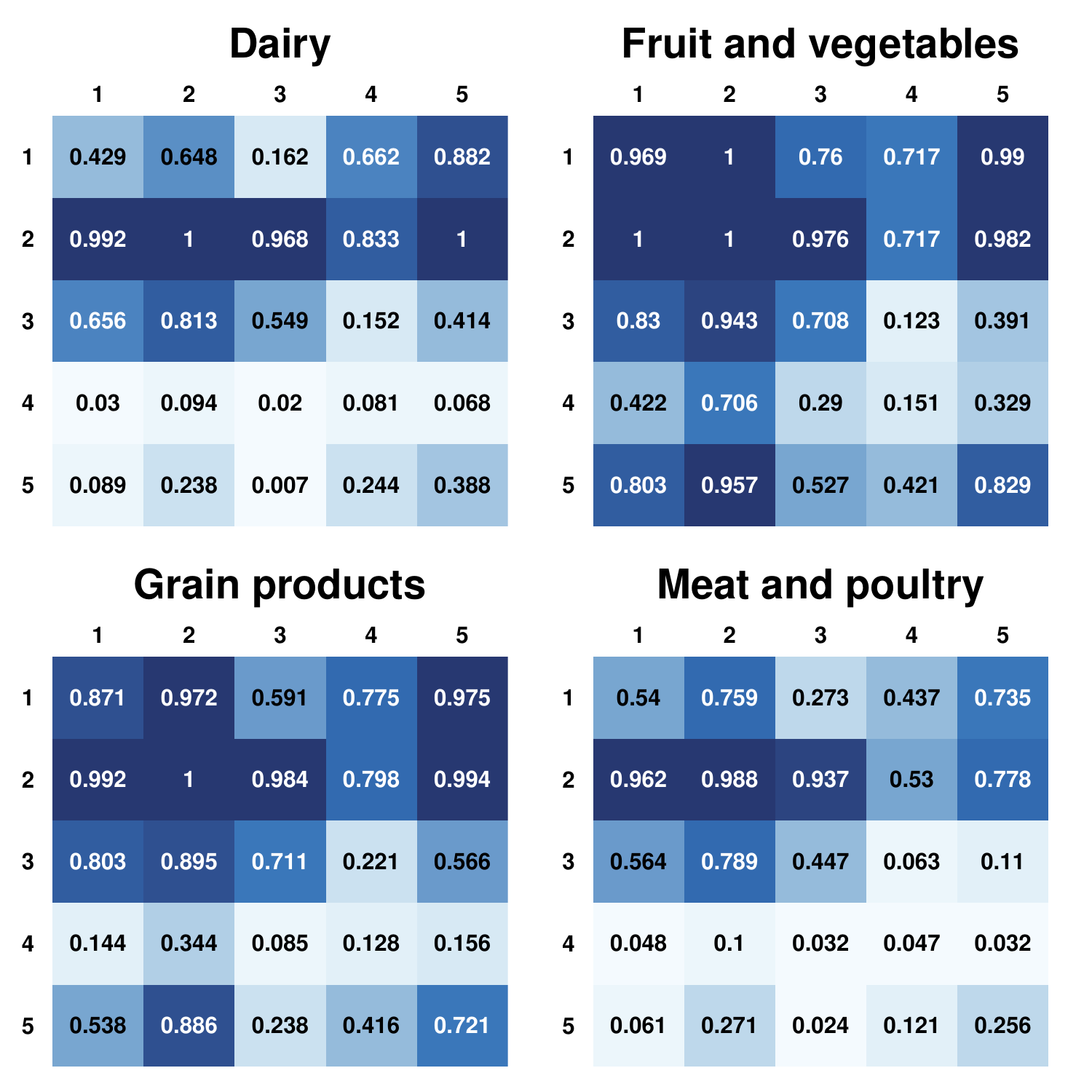}
\subcaption{}
\label{fao_data_p_K=5}
\end{subfigure}
\begin{subfigure}{1\textwidth}
\centering
\includegraphics[width=0.75\linewidth,valign=t]{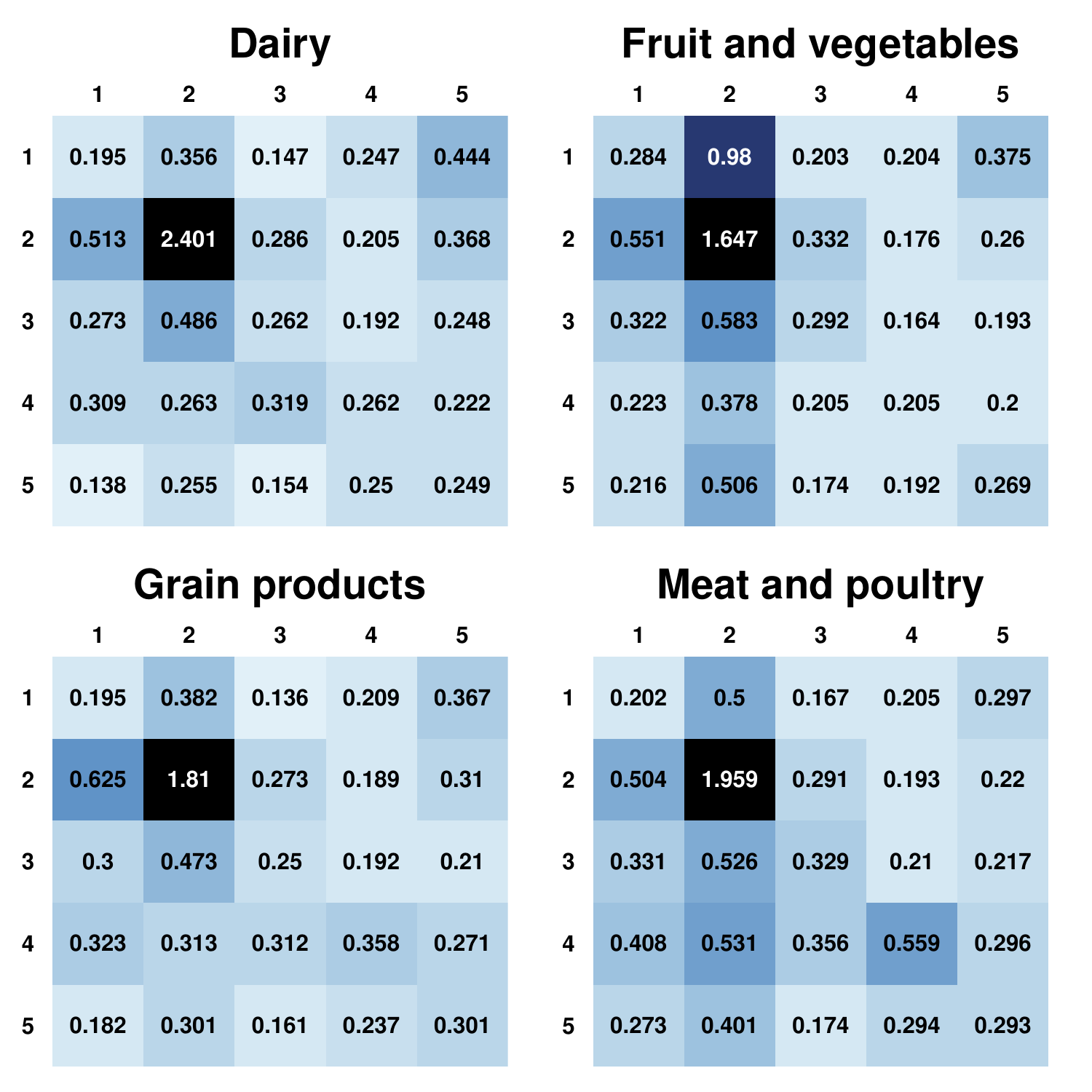}
\subcaption{}
\label{fao_data_alpha_K=5}
\end{subfigure}
\caption{Parameter estimates of multi-DirSBM with 5 clusters on FAO trade data for the 4 product categories considered: \subref{fao_data_p_K=5} estimated connectivity matrices $\mathcal{P}$; \subref{fao_data_alpha_K=5} estimated Dirichlet concentration matrices $\mathcal{A}$. Rows correspond to the sending cluster, while columns correspond to the receiving cluster. Darker colors indicate higher values of parameter estimates.}
\label{fao_data_params}
\end{figure}

The map of the world presented in Figure \ref{fao_data_map_K=5} reports the clustering of the countries returned by the selected model. A complete list of countries allocated to their respective clusters is given in Appendix \ref{list_of_countries}. Figure \ref{fao_data_cluster_shares_K=5} illustrate the percentage flows between clusters in each of the 4 product categories considered, as defined in Equation \eqref{V_s}. In this figure, the labels S1 to S5 on the left correspond to the sending cluster label (1 to 5, respectively), with the percentage flows summing to 100\% for each the sending cluster. The labels R1 to R5 correspond to the receiving cluster label. Figure \ref{fao_data_p_K=5} displays heatmaps of the estimated connectivity matrices in the four product categories, while Figure \ref{fao_data_alpha_K=5} shows heatmaps of the estimated Dirichlet concentration parameters in each of the product categories. 

Examining Figure \ref{fao_data_map_K=5}, we note that the clusters are to some degree associated with the geographical location and the economic development of the countries. The orange cluster (cluster 1) comprises medium-sized European economies, such as Austria and Norway, along with Turkey and Israel. The purple cluster (cluster 2) is the smallest, with only 10 countries and it contains some of the largest world economies, such as the US, Germany, France, Russia, China, all located in the Northern hemisphere. The blue cluster (cluster 3) primarily includes countries in Eastern and Northern Europe as well as the Balkan Peninsula, and it consists of smaller European economies, such as Latvia, Macedonia and Serbia. This cluster also contains Cyprus and Luxembourg.  The red cluster (cluster 4) is the most populous, including 22 countries. It comprises all of the South American economies considered, such as Brazil and Argentina, large African economies (e.g. Ghana, Kenya, Tunisia) as well as some Middle Eastern countries (e.g. Iran, Afghanistan, Pakistan) and Sri Lanka. Lastly, the green cluster (cluster 5) is the one with the largest geographical spread, comprising large and medium economies across Asia (e.g. India, Indonesia, South Korea, countries in the Arabian Peninsula), Africa (e.g. Morocco and South Africa), as well as Canada, Australia, New Zealand and Mexico.

According to Figure \ref{fao_data_p_K=5}, countries in cluster 1 (orange) have consistently high estimated probabilities of exporting products from all four categories to countries in cluster 5 (green) and cluster 2 (purple). Moreover, the associated trades represent large relative volumes, as indicated by the estimated Dirichlet concentration parameters in Figure \ref{fao_data_alpha_K=5}. Notably, the fruit and vegetable exports of the orange countries to purple countries (cluster 2) are of very large relative volumes, accounting for over 46\% of cluster 1's total fruit and vegetable exports. For cluster 1, the fruit and vegetables product category corresponds to the highest estimated intra-cluster connectivity, 0.969, as reported in Figure \ref{fao_data_p_K=5}, indicating that countries in this cluster tend to frequently exchange these products. Relatively low estimated probabilities of connection observed in all food categories are to the red countries (cluster 3), along with low Dirichlet concentration parameter estimates in Figure \ref{fao_data_alpha_K=5}. In contrast, the relatively low estimated probabilities of connection to the red countries (cluster 4) across all food categories, along with low Dirichlet concentration parameter estimates in Figure \ref{fao_data_alpha_K=5}, suggest that exports from orange countries to red countries are less frequent and involve relatively low volumes.

From Figure \ref{fao_data_p_K=5}, cluster 2 (purple) is the most connected cluster, with high estimated connectivity probability to all other clusters across all product categories. The cluster itself is highly interconnected, with three out of four estimated within cluster connectivity probabilities in Figure \ref{fao_data_p_K=5} being numerically equal to one. Trade between its members across all product categories represents the largest relative volumes, as the Dirichlet concentration parameter estimates in Figure \ref{fao_data_alpha_K=5} tend to be remarkably higher than those corresponding to trades to other clusters. This is reflected in the overall cluster trade shares in Figure \ref{fao_data_cluster_shares_K=5}, where anywhere between 41\% and 52\% of the purple cluster's export volumes are traded within the cluster. Additionally, purple countries tend to export extensively to the orange (cluster 1) countries, although in smaller percentages as the Dirichlet concentration parameter estimates in Figure \ref{fao_data_alpha_K=5} suggest. 

Cluster 3 (blue) shows the highest connectivity probabilities across all the categories with cluster 2 (purple) countries, and its exports to purple countries represent the largest relative volumes, as shown in figures \ref{fao_data_p_K=5} and \ref{fao_data_alpha_K=5}, respectively. Furthermore, exports from the blue cluster to the purple cluster account for the largest total cluster-to-cluster export shares, ranging from 35\% of grain exports to 44\% of meat and poultry exports.  The lowest connectivity probabilities are observed with cluster 4 (red) countries in all the product categories, and the exports to red countries also seem to be of the lowest relative volumes, according to Figure \ref{fao_data_alpha_K=5}. These limited connections and small relative export volumes result in the smallest expected cluster-to-cluster export shares, with less than 5\% of blue cluster's exports in all the food categories being to the red cluster (cluster 4).

Looking at cluster 4 (red), its members appears to have limited export connections in most product categories compared to other clusters, as the estimated connectivity probabilities are consistently among the lowest in each category, according to Figure \ref{fao_data_p_K=5}. Notably, countries in the red cluster are likely to have particularly sparse export connections in dairy and meat and poultry products, suggesting that these two categories are perhaps not their most important exports. Nonetheless, out of its total meat and poultry trades, the red countries seem to trade relatively large shares with other red countries as well as purple (cluster 2) countries, as indicated by relatively large Dirichlet parameter estimates in Figure \ref{fao_data_alpha_K=5}. By total volume, according to Figure \ref{fao_data_cluster_shares_K=5}, the largest shares of dairy and meat and poultry exports are traded within the cluster itself (43\% and 35\% respectively). Higher connectivity is observed in the red cluster's fruit and vegetable export network, with the highest estimated connectivity probability of 0.706 to the purple (cluster 2) countries. When combined, fruit and vegetables exports to the purple cluster (cluster 2) account for 32\% of red cluster's fruit and vegetables trade.

Lastly, considering cluster 5 (green), its connectivity in dairy and meat and poultry appears to be limited, as indicated by the connectivity probability parameter estimates in Figure \ref{fao_data_p_K=5}. However, for fruit and vegetables, the green countries have high estimated probabilities of exporting to purple (cluster 2), green (cluster 5), and orange (cluster 1) countries. The exports to cluster 2 countries are also of relatively large volumes based on the Dirichlet parameter estimates from Figure \ref{fao_data_alpha_K=5}. Furthermore, fruit and vegetable exports to cluster 2 account for over 31\% of green cluster's total fruit and vegetables exports, with an additional 29\% traded within the cluster itself. With regards to grain products, the highest estimated connectivity probabilities are observed to purple countries (cluster 2) and other green countries (cluster 5). The relative shares exported to countries in different clusters are fairly homogeneous as the Dirichlet concentration parameter estimates in Figure \ref{fao_data_alpha_K=5} tend to be similar across clusters. Thus, the differences in total cluster-to-cluster export shares in Figure \ref{fao_data_cluster_shares_K=5} are largely due to the number of connections cluster 5 has with each cluster and the number of countries in each respective cluster.

In summary, the analysis of the clusters inferred by multi-DirSBM reveals distinct trade and connectivity patterns across the five identified clusters. Cluster 1 (orange) exhibits high export connectivity with clusters 5 (green) and 2 (purple), particularly in fruit and vegetable exports, while maintaining lower trade volumes with cluster 4 (red). Cluster 2 (purple) is the most interconnected, with high connectivity probabilities across all clusters and large intra-cluster trade volumes, in all food categories. Cluster 3 (blue) has trade relationship primarily with cluster 2, particularly in grain and meat exports. Cluster 4 (red) has sparse connectivity, especially in trading dairy and meat products, with stronger internal trade and some exchange with purple countries. Cluster 5 (green) exhibits diverse export connections, especially in fruit and vegetables, with substantial trade with both purple and orange countries, and a more uniform distribution of grain exports. Overall, the results highlight the regional and economic specificity of trade flows, with some clusters displaying both high intra-cluster cohesion and substantial external trade relations.

\section{Discussion} \label{Discussion}

This paper proposed a multiplex Dirichlet stochastic block model, multi-DirSBM, a novel approach to clustering multidimensional compositional networks. The approach is based on two key contributions, which extend the standard Dirichlet stochastic block model, DirSBM, developed to model fully connected composition-weighted single-layer networks. First, it incorporates a Bernoulli distribution to model the presence of edges between node pairs, enabling modeling compositional networks with absent edges and aligning with common practices in stochastic block models. Second, it generalizes the model to multiplex networks, enabling joint clustering of different types of interactions. Estimation of the proposed multi-DirSBM is based on a hybrid likelihood and a variant of the classification EM algorithm. Model selection is performed using ICL and BIC criteria. A simulation study is carried to assess the model's performance, which has shown satisfactory results in terms of clustering, parameter estimation, and model selection, highlighting the advantages of modeling both the multidimensional network structure and compositional edges. Finally, the multi-DirSBM has been applied to analyse the multiplex FAO trade network, revealing insightful clustering patterns of major food-exporting nations.

A number of future directions and developments of this work are possible. First of all, similarly to the original DirSBM \citep{dir_sbm}, the proportional data are considered with respect to the sending nodes, hence the resulting partition reflects the patterns associated with the proportions being sent by the nodes. Alternatively, rather than defining compositional vectors in terms of sending nodes, one could compute compositions with respect to receiving nodes. For example, in relation to the FAO data, an extension of this approach could involve examining relative import volumes, where the compositional weights represent a country's share of total imports in a product category, rather than exports. Additionally, both sending and receiving proportion data could be combined into a single multiplex network, with layers denoting product categories and exports/imports flows. This would allow for the analysis of the combined effect of export and import proportions on the FAO network's latent structure. However, caution may be needed when combining import and export compositional relational weights.

An additional avenue of future research consists in the exploration of more efficient implementations of the estimation algorithm. The current version is only suitable for networks of moderate size and/or density due to the dimensionality of the Dirichlet-distributed edge weights, which is tied to both the density of the network and the number of nodes. Future implementations could focus on improving the speed of the Dirichlet concentration parameter estimation. For instance, considering approximations of the Dirichlet distribution that are more convenient to work with could help streamline this process \citep{dir_approx}.

Another future direction could involve extending the multi-DirSBM to temporal networks, where individual layers of the multiplex represent different time points. Here, we have only considered a snapshot of the FAO network from 2010. However, many network datasets, including FAO trade data, are collected over time, it would be valuable to account for the dynamic nature of such data. This would to uncover clustering patterns and examine how these patterns evolve over time. This could be achieved by introducing dependence between layers through cluster labels that change over time. A more general extension of multi-DirSBM to temporal networks would also allow the model parameters to evolve across different time points \citep{dynamic_sbm,po_dc_dynamic_sbm}.

\newpage
\appendix
\section{Classification EM algorithm for multi-DirSBM} \label{app_CEM_steps}

Presented here are the derivations of the classification EM algorithm steps for multi-DirSBM. The derivations of steps for DirSBM with absent edges can be recovered by simply setting the number of layers equal to 1, i.e. $S=1$.

Using a shorthand notation $\mathbf{e}_i^{(1:S)}=\{\mathbf{e}_i^{(1)},\ldots,\mathbf{e}_i^{(S)}\}$ and $\mathbf{x}_i^{(1:S)}=\{\mathbf{x}_i^{(1)},\ldots,\mathbf{x}_i^{(S)}\}$, the E-step proceeds in the standard EM algorithm fashion by making use of Bayes' rule:

\begin{equation*}
\begin{aligned}
    \hat{z}_{ik} & = p(z_{ik}=1|\mathbf{e}_i^{(1:S)},\mathbf{x}_i^{(1:S)},\widetilde{\mathbf{Z}}_{-i})\\[1.5ex] & =\dfrac{p(z_{ik}=1|\widetilde{\mathbf{Z}}_{-i})p(\mathbf{e}_i^{(1:S)},\mathbf{x}_i^{(1:S)}|z_{ik}=1,\widetilde{\mathbf{Z}}_{-i})}{p(\mathbf{e}_i^{(1:S)},\mathbf{x}_i^{(1:S)}|\widetilde{\mathbf{Z}}_{-i})} \\[1.5ex] &  =\dfrac{p(z_{ik}=1)\prod_{s=1}^{S}p(\mathbf{x}_i^{(s)}|\mathbf{e}_i^{(s)},z_{ik}=1,\widetilde{\mathbf{Z}}_{-i})p(\mathbf{e}_i^{(s)}|z_{ik}=1,\widetilde{\mathbf{Z}}_{-i})}{\sum_{h=1}^{K}p(z_{ih}=1) \prod_{s=1}^{S} p(\mathbf{x}_i^{(s)}|\mathbf{e}_i^{(s)},z_{ih}=1,\widetilde{\mathbf{Z}}_{-i})p(\mathbf{e}_i^{(s)}|z_{ih}=1,\widetilde{\mathbf{Z}}_{-i})} \\[1.5ex] & 
    \propto p(z_{ik}=1)^{(s)} p(\mathbf{x}_i^{(s)}|\mathbf{e}_i^{(s)},z_{ik}=1,\widetilde{\mathbf{Z}}_{-i})p(\mathbf{e}_i^{(s)}|z_{ik}=1,\widetilde{\mathbf{Z}}_{-i}) \\[1.5ex]& =
    \theta_k \left[\prod_{s=1}^{S}\prod_{j=1}^{n}\prod_{h=1}^{K} \left[(1-p_{kh}^{(s)})^{1-e_{ij}^{(s)}}p_{kh}^{(s)e_{ij}^{(s)}}\right]^{\Tilde{z}_{jh}}\right]\left[\frac{\Gamma(\sum_{j=1}^{n} e_{ij} \Tilde{\alpha}_{j}^{(s)})}{\prod_{j=1}^{n}\Gamma(\Tilde{\alpha}_{j}^{(s)})^{e_{ij}^{(s)}}} \prod_{j=1}^{n} x_{ij}^{(s)e_{ij}^{(s)} (\Tilde{\alpha}_{j}^{(s)}-1)}\right]
    \end{aligned}
\end{equation*}

The expectation of the complete data hybrid log-likelihood $l_{c}^{hyb}$ from Equation \eqref{complete_hybrid_ll_multi} with respect to a single latent variable $\mathbf{z}_i$ is given by

\begin{equation} \label{E_of_complete_hybrid_ll_multi}
\begin{aligned}
    \mathbb{E}[l_{c}^{hyb}] = \sum_{s=1}^{S} \sum_{i=1}^{n} \sum_{k=1}^{K} \mathbb{E}[z_{ik}] \Bigg( &
    \log \theta_k + \sum_{j=1}^{n} \sum_{h=1}^{K} \Tilde{z}_{jh} (1-e_{ij}^{(s)}) \log (1-p_{kh}^{(s)}) + \sum_{j=1}^{n} \sum_{h=1}^{K} \Tilde{z}_{jh} e_{ij}^{(s)} \log p_{kh}^{(s)} \\ & + \log \Gamma (\sum_{j=1}^{n} e_{ij}^{(s)} \Tilde{\alpha}_{j}^{(s)} +\mathbb{I}\{\sum_{j=1}^{n}e_{ij}^{(s)}=0\}) - \sum_{j=1}^{n} e_{ij}^{(s)} \log \Gamma (\Tilde{\alpha}_{j}^{(s)}) \\& +\sum_{j=1}^{n} e_{ij}^{(s)} (\Tilde{\alpha}_{j}^{(s)} - 1) \log x_{ij}^{(s)} \Bigg).
    \end{aligned}
\end{equation}

\noindent This is the objective function maximised in the M-step with respect to the mixing proportions $\boldsymbol{\theta}$, the set of connectivity matrices $\mathcal{P}$ and the set of Dirichlet parameter matrices $\mathcal{A}$. 

The updates for the mixing proportions $\theta_k$ are available in closed form and are derived in the same way as for the original DirSBM by maximising $\mathbb{E}[l_c^{hyb}]$, subject to constraint $\sum_{k} \theta_k = 1$. Let

\begin{equation}
    f = \mathbb{E}[l_c^{hyb}] - \lambda (\sum_{k=1}^{K} \theta_k - 1).
\end{equation}

The partial derivative of $f$ with respect to $\theta_k$ is

\begin{equation*}
    \dfrac{1}{\theta_k}\sum_{i=1}^{n} \hat{z}_{ik} -\lambda = 0,
\end{equation*}

\noindent which gives $\lambda \theta_k = \sum_{i=1}^{n} \hat{z}_{ik}$. Summing over k and using the unit-sum constraint for $(\theta_1,...,\theta_K)$, we find that $\lambda=n$, producing the update for the mixing proportions from Equation \eqref{theta-m-step_multi}. 

The update for the set of connectivity matrices $\mathcal{P}$ are found layer-wise and entry-wise by solving 

\begin{equation*}
\begin{aligned}
    & \frac{\partial \mathbb{E}[l_c^{hyb}]}{\partial p_{kh}^{(s)}} = 
    \sum_{i=1}^{n} \hat{z}_{ik} \left\{ - \frac{1}{1-p_{kh}^{(s)}} \sum_{j=1}^{n} \Tilde{z}_{jh} (1-e_{ij}^{(s)}) + \frac{1}{p_{kh}^{(s)}} \sum_{j=1}^{n} \Tilde{z}_{jh} e_{ij}^{(s)} \right\} = 0
    \\[1ex]
    & \sum_{i=1}^{n} \sum_{j=1}^{n} \left\{ -  \hat{z}_{ik} \Tilde{z}_{jh} p_{kh}^{(s)} +  \cancel{\hat{z}_{ik} \Tilde{z}_{jh} e_{ij}^{(s)} p_{kh}^{(s)}} +  \hat{z}_{ik} \Tilde{z}_{jh} e_{ij}^{(s)} - \cancel{ \hat{z}_{ik} \Tilde{z}_{jh} e_{ij}^{(s)} p_{kh}^{(s)}} \right\} = 0
    \\[1ex]
    & p_{kh}^{(s)} \sum_{i=1}^{n} \sum_{j=1}^{n} \hat{z}_{ik} \Tilde{z}_{jh}  = \sum_{i=1}^{n} \sum_{j=1}^{n} \hat{z}_{ik} \Tilde{z}_{jh} e_{ij}^{(s)}. 
\end{aligned}
\end{equation*}

This leads to the closed form update in Equation \eqref{P-m-step_multi}.

\section{Model selection: Integrated completed likelihood and Bayesian information criterion} \label{app_icl}

The derivation of the integrated completed likelihood (ICL) criterion for the multi-DirSBM follows closely the derivation for the original DirSBM.

Following \cite{ICL}, the exact complete-data integrated log-likelihood is given by

\begin{equation} \label{exact_ICL}
    \log p(\mathcal{X},\mathcal{E},\mathbf{Z}|K) = \log p(\mathcal{X},\mathcal{E}|\mathbf{Z},K) + \log p(\mathbf{Z}|K).
\end{equation}

Following closely the derivations of the ICL for random graphs in \cite{Daudin_mm_random_graphs}, we can find the first term of Equation \eqref{exact_ICL} using a BIC approximation:

\begin{equation} \label{bic_approx}
\begin{aligned}
    \log p(\mathcal{X},\mathcal{E} \lvert \mathbf{Z},K) = & \max_{\mathcal{P},\mathcal{A}} \log p(\mathcal{X},\mathcal{E} \lvert \mathbf{Z},\mathcal{P},\mathcal{A},K) \\ & - \frac{1}{2} 2 K^2 S \log S n(n-1) \\ &
    \hspace{-3mm} = \max_{\mathcal{P},\mathcal{A}} \log p(\mathcal{X},\mathcal{E} \lvert \mathbf{Z},\mathcal{P},\mathcal{A},K) \\ & - K^2 S \log S n(n-1) ,
\end{aligned}
\end{equation}
where $2 K^2 S = (K^2 + K^2)S$ is the number of elements of matrix $\mathbf{A}^{(s)}$ plus the number of elements of $\mathbf{P}^{(s)}$ in every layer, and $S n(n-1)$ is the number of all possible edges in the network (that excludes self-loops.

The second term of Equation \eqref{exact_ICL} is derived in the same fashion as in \cite{Daudin_mm_random_graphs}, using a non-informative Jeffrey prior and Stirling formula for an approximation of a Gamma function, leading to

\begin{equation}
    \log p(\mathbf{Z}|K) = \max_{\boldsymbol{\theta}} \log p(\mathbf{Z}|\boldsymbol{\theta},K) - \frac{1}{2} (K-1) \log n,
\end{equation}
with $(K-1)$ being the number of free parameters in $\boldsymbol{\theta}$ and $n$ being the number of latent variables in the model.

Substituting the approximation results back and then replacing the complete data log-likelihood with its hybrid counterpart in the fashion of \cite{HybridML}, we arrive at 

\begin{equation}
\begin{aligned}
    ICL(K) & = \log p(\mathcal{X},\mathcal{E}|\mathbf{\hat{Z}},\hat{\mathcal{P}},\hat{\mathcal{A}},K) - K^2 S \log S n(n-1) \\& \hspace{4mm} + \log p (\mathbf{\hat{Z}}|\boldsymbol{\hat{\theta}},K) -\frac{1}{2}(K-1)\log n \\ 
    & = \log p(\mathcal{X},\mathcal{E},\mathbf{\hat{Z}}|\hat{\mathcal{P}},\hat{\mathcal{A}},\boldsymbol{\hat{\theta}},K) - K^2 S \log S n(n-1) -\frac{1}{2}(K-1)\log n \\
    & = l^{hyb}_c(\hat{\mathcal{P}},\hat{\mathcal{A}},\boldsymbol{\hat{\theta}}|\mathcal{X},\mathcal{E},\mathbf{\hat{Z}}) - K^2 S \log S n(n-1) -\frac{1}{2}(K-1)\log n.
\end{aligned}
\end{equation}

Bayes information criterion used for model selection in this work can be obtained from Equation \eqref{bic_approx}, by replacing the maximum of the observed data log-likelihood $\max_{\mathcal{P},\mathcal{A}} \log p(\mathcal{X},\mathcal{E} \lvert \mathbf{Z},\mathcal{P},\mathcal{A},K)$ by the hybrid log-likelihood with the parameter estimates and the estimated partition plugged in, i.e. $\log p(\mathcal{X},\mathcal{E}|\mathbf{\hat{Z}},\hat{\mathcal{P}},\hat{\mathcal{A}},K)$.

Note that in order to derive a model selection criterion for DirSBM for compositional networks with zeros, one is only required to set $S=1$ in the above derivations.

\section{Alternative data generating procedure} \label{app_data_gen_gamma}

Similarly to the original DirSBM by \cite{dir_sbm}, an alternative data generating process for multi-DirSBM can be formulated using independent Gamma samples:

\begin{enumerate}
    \item Given a $K$-dimensional vector of mixing proportions $\boldsymbol{\theta}$, assign each node $i$ to cluster $k$ with probability $\theta_k$, and denote by $\mathbf{c}$ the resulting cluster assignments vector. Note that the partition into clusters is assumed to be constant across the layers of the network.

    \item Let $S$ be the number of layers in a multiplex network. Let $\mathcal{P}=\{\mathbf{P}^{(1)},\ldots,\mathbf{P}^{(s)}, \ldots, \mathbf{P}^{(S)}\}$ be a collection of $K\times K$ matrices of connectivity probabilities in each of the $S$ layers. Generate binary edges $e_{ij}^{(s)}$ between nodes $i$ and $j$ in layer $s$ using 
    $$
        e_{ij}^{(s)}|c_i=k,c_j=h \sim Bernoulli(p_{kh}^{(s)}).
    $$
    
    and store them in the set of adjacency matrices $\mathcal{E}=\{\mathbf{E}^{(1)},\ldots,\mathbf{E}^{(s)},\ldots,\mathbf{E}^{(S)}\}$.
     \item Given a collection of $K \times K$ parameter matrices with strictly positive values $\mathcal{A}=\{\mathbf{A}^{(1)},\ldots,\mathbf{A}^{(s)},\ldots,\mathbf{A}^{(S)}\}$, for pairs of nodes $i$ and $j$ in layer $s$, independent weights are generated from
    $$
    y_{ij}^{(s)} \lvert e_{ij}=1,c_i=k,c_j=h \sim Gamma(\alpha_{kh}^{(s)},1).
    $$
    When $e_{ij}=1$, the associated edge weight $y_{ij}^{(s)}=0$.
    \item For each sending node $i$, construct compositional weights vector $\mathbf{x}_{i}^{(s)}$ from the collection of independent weights:
    $$
    \mathbf{x}_{i}^{(s)}= \bigg(\frac{y_{i1}^{(s)}}{\sum_{j\neq i}^{n}y_{ij}^{(s)}},\ldots,\frac{y_{in}^{(s)}}{\sum_{j\neq i}^{n}y_{ij}^{(s)}} \bigg).
    $$
\end{enumerate}

The version of $\mathbf{x}_{i}^{(s)}$ without zero entries is equivalent to $\mathbf{x}_{i}^{(s)*}$ in step 3 of the data generating procedure described in Section \ref{Extensions} in the main text. 

\section{Simulation studies: supplementary notes} \label{app_sim_params}

The Dirichlet concentration parameter matrices used in the simulation studies are provided below. For scenarios with multiplex data with two layers, only the first two matrices $\mathbf{A}_1$ and $\mathbf{A}_2$ have been used to generate the data.

\vspace{3mm}
\noindent $K=2$
\begin{equation}
    \mathbf{A}^{(1)}= \begin{pmatrix} 1.0 & 0.6 \\ 0.9 & 1.4 \end{pmatrix} \hspace{5mm} 
    \mathbf{A}^{(2)}= \begin{pmatrix} 1.2 & 0.7 \\ 0.5 & 1 \end{pmatrix} \hspace{5mm} 
    \mathbf{A}^{(3)}= \begin{pmatrix} 1.1 & 0.5 \\ 0.4 & 0.9 \end{pmatrix} \hspace{5mm}
    \mathbf{A}^{(4)}= \begin{pmatrix} 1.3 & 0.6 \\ 0.8 & 1.2 \end{pmatrix};
\end{equation}

\vspace{3mm}
\noindent $K=3$
\begin{equation}
    \mathbf{A}^{(1)}= \begin{pmatrix} 1.0 & 0.7 & 0.5 \\ 0.9 & 1.5 & 0.6 \\ 0.4 & 0.5 & 1.2  \end{pmatrix} \hspace{3mm}   
    \mathbf{A}^{(2)}= \begin{pmatrix} 1.1 & 0.7 & 0.4 \\ 0.8 & 1 & 0.5 \\ 0.4 & 0.6 & 1.3  \end{pmatrix} \hspace{3mm} 
\end{equation}
\begin{equation}
    \mathbf{A}^{(3)}= \begin{pmatrix} 1.0 & 0.5 & 0.7 \\ 0.9 & 1.2 & 0.6 \\ 0.3 & 0.5 & 0.9  \end{pmatrix} \hspace{3mm} 
    \mathbf{A}^{(4)}= \begin{pmatrix} 1.3 & 0.8 & 0.7 \\ 0.6 & 1.1 & 0.5 \\ 0.7 & 0.4 & 1.2  \end{pmatrix};
\end{equation}

\vspace{3mm}
\noindent $K=5$
\begin{equation}
    \mathbf{A}^{(1)}= \begin{pmatrix} 1.0 & 0.7 & 0.5 & 0.4 & 0.6 \\ 0.9 & 1.5 & 0.6 & 0.5 & 0.7 \\ 0.4 & 0.5 & 1.2 & 0.6 & 0.3 \\ 0.8 & 0.6 & 0.4 & 1.4 & 0.5 \\ 0.5 & 0.8 & 0.9 & 0.7 & 1.7  \end{pmatrix} \hspace{5mm}
    \mathbf{A}^{(2)}= \begin{pmatrix} 1.1 & 0.7 & 0.4 & 0.5 & 0.6 \\ 0.8 & 1 & 0.5 & 0.4 & 0.7 \\ 0.4 & 0.6 & 1.3 & 0.6 & 0.3 \\ 0.9 & 0.8 & 0.6 & 1.4 & 0.5 \\ 0.5 & 0.7 & 0.9 & 0.8 & 1.7  \end{pmatrix} 
\end{equation}

\begin{equation}    
    \mathbf{A}^{(3)}= \begin{pmatrix} 1.0 & 0.7 & 0.5 & 0.6 & 0.4 \\ 0.9 & 1.2 & 0.6 & 0.5 & 0.7 \\ 0.3 & 0.5 & 0.9 & 0.4 & 0.6 \\ 0.8 & 0.6 & 0.4 & 1.5 & 0.5 \\ 0.5 & 0.9 & 0.8 & 0.7 & 1.4  \end{pmatrix} \hspace{5mm}
    \mathbf{A}^{(4)}= \begin{pmatrix} 1.3 & 0.8 & 0.7 & 0.4 & 0.6 \\ 0.6 & 1.1 & 0.8 & 0.5 & 0.7 \\ 0.7 & 0.4 & 1.2 & 0.6 & 0.3 \\ 0.8 & 0.6 & 0.4 & 1.4 & 0.5 \\ 0.9 & 0.7 & 0.5 & 0.8 & 1.5  \end{pmatrix}.
\end{equation}

\vspace{5mm}
\noindent The connectivity parameter matrices producing networks with different densities and cluster overlaps are given below.

\vspace{3mm}
\noindent $K=2$, $S=2$, high density, no cluster overlap
\begin{equation}
    \mathbf{P}^{(1)}= \begin{pmatrix} 0.9 & 0.5 \\ 0.3 & 0.8 \end{pmatrix} \hspace{5mm} 
    \mathbf{P}^{(2)}= \begin{pmatrix} 0.8 & 0.6 \\ 0.5 & 0.7 \end{pmatrix};
\end{equation}

\vspace{3mm}
\noindent $K=2$, $S=4$, low density, some cluster overlap
\begin{equation}
    \mathbf{P}^{(1)}= \begin{pmatrix} 0.5 & 0.3 \\ 0.4 & 0.6 \end{pmatrix} \hspace{3mm} 
    \mathbf{P}^{(2)}= \begin{pmatrix} 0.45 & 0.3 \\ 0.35 & 0.55 \end{pmatrix} \hspace{3mm} 
    \mathbf{P}^{(3)}= \begin{pmatrix} 0.5 & 0.4 \\ 0.3 & 0.45 \end{pmatrix} \hspace{3mm}
    \mathbf{P}^{(4)}= \begin{pmatrix} 0.45 & 0.35 \\ 0.4 & 0.6 \end{pmatrix};
\end{equation}

\vspace{3mm}
\noindent $K=3$, $S=2$, high density, some cluster overlap
\begin{equation}
    \mathbf{P}^{(1)}= \begin{pmatrix} 0.9 & 0.5 & 0.75 \\ 0.45 & 0.8 & 0.65 \\ 0.6 & 0.55 & 0.7 \end{pmatrix} \hspace{5mm}   
    \mathbf{P}^{(2)}= \begin{pmatrix} 0.75 & 0.5 & 0.6 \\ 0.65 & 0.85 & 0.55 \\ 0.8 & 0.6 & 0.9  \end{pmatrix};
\end{equation}

\vspace{3mm}
\noindent $K=3$, $S=2$, high density, no cluster overlap
\begin{equation}
    \mathbf{P}_1= \begin{pmatrix} 0.9 & 0.5 & 0.4 \\ 0.3 & 0.6 & 0.1 \\ 0.5 & 0.2 & 0.7  \end{pmatrix} \hspace{5mm}   
    \mathbf{P}_2= \begin{pmatrix} 0.75 & 0.3 & 0.2 \\ 0.6 & 0.85 & 0.1 \\ 0.5 & 0.6 & 0.9  \end{pmatrix};
\end{equation}

\vspace{3mm}
\noindent $K=3$, $S=4$, low density, no cluster overlap
\begin{equation}
    \mathbf{P}^{(1)}= \begin{pmatrix} 0.5 & 0.1 & 0.3 \\ 0.2 & 0.4 & 0.15 \\ 0.4 & 0.3 & 0.6  \end{pmatrix} \hspace{3mm}   
    \mathbf{P}^{(2)}= \begin{pmatrix} 0.45 & 0.2 & 0.1 \\ 0.35 & 0.6 & 0.4 \\ 0.15 & 0.3 & 0.55  \end{pmatrix} \hspace{3mm} 
\end{equation}
\begin{equation}
    \mathbf{P}^{(3)}= \begin{pmatrix} 0.45 & 0.3 & 0.2 \\ 0.2 & 0.35 & 0.1 \\ 0.3 & 0.1 & 0.5  \end{pmatrix} \hspace{3mm} 
    \mathbf{P}^{(4)}= \begin{pmatrix} 0.7 & 0.4 & 0.3 \\ 0.35 & 0.5 & 0.2 \\ 0.1 & 0.15 & 0.4  \end{pmatrix};
\end{equation}

\vspace{3mm}
\noindent $K=3$, $S=4$, high density, some cluster overlap
\begin{equation}
    \mathbf{P}^{(1)}= \begin{pmatrix} 0.7 & 0.4 & 0.5 \\ 0.4 & 0.8 & 0.6 \\ 0.5 & 0.6 & 0.85  \end{pmatrix} \hspace{3mm}   
    \mathbf{P}^{(2)}= \begin{pmatrix} 0.9 & 0.6 & 0.7 \\ 0.5 & 0.7 & 0.4 \\ 0.7 & 0.5 & 0.8  \end{pmatrix} \hspace{3mm} 
\end{equation}
\begin{equation}
    \mathbf{P}^{(3)}= \begin{pmatrix} 0.9 & 0.5 & 0.7 \\ 0.6 & 0.75 & 0.4 \\ 0.5 & 0.6 & 0.8  \end{pmatrix} \hspace{3mm} 
    \mathbf{P}^{(4)}= \begin{pmatrix} 0.75 & 0.5 & 0.6 \\ 0.6 & 0.85 & 0.7 \\ 0.5 & 0.65 & 0.8  \end{pmatrix};
\end{equation}

\vspace{3mm}
\noindent $K=5$, $S=2$, high density, no cluster overlap
\begin{equation}
    \mathbf{P}^{(1)}= \begin{pmatrix} 0.7 & 0.4 & 0.2 & 0.5 & 0.3 \\ 0.4 & 0.8 & 0.3 & 0.6 & 0.2 \\ 0.5 & 0.2 & 0.85 & 0.1 & 0.4 \\ 0.6 & 0.3 & 0.4 & 0.9 & 0.5 \\ 0.1 & 0.6 & 0.5 & 0.2 & 0.75  \end{pmatrix} \hspace{5mm}
    \mathbf{P}^{(2)}= \begin{pmatrix} 0.9 & 0.4 & 0.2 & 0.6 & 0.1 \\ 0.3 & 0.7 & 0.1 & 0.5 & 0.4 \\ 0.5 & 0.2 & 0.75 & 0.1 & 0.3 \\ 0.6 & 0.1 & 0.3 & 0.85 & 0.5 \\ 0.2 & 0.5 & 0.4 & 0.3 & 0.8  \end{pmatrix};
\end{equation}

\vspace{3mm}
\noindent $K=5$, $S=4$, low density, some cluster overlap
\begin{equation}
    \mathbf{P}^{(1)}= \begin{pmatrix} 0.5 & 0.1 & 0.3 & 0.2 & 0.4 \\ 0.2 & 0.4 & 0.15 & 0.25 & 0.1 \\ 0.1 & 0.2 & 0.45 & 0.3 & 0.35 \\ 0.3 & 0.15 & 0.4 & 0.6 & 0.2 \\ 0.4 & 0.3 & 0.2 & 0.15 & 0.55  \end{pmatrix} \hspace{5mm}
    \mathbf{P}^{(2)}= \begin{pmatrix} 0.45 & 0.3 & 0.1 & 0.35 & 0.25 \\ 0.4 & 0.55 & 0.3 & 0.25 & 0.1 \\ 0.35 & 0.45 & 0.6 & 0.3 & 0.25 \\ 0.3 & 0.15 & 0.25 & 0.5 & 0.35 \\ 0.1 & 0.35 & 0.2 & 0.15 & 0.4  \end{pmatrix} 
\end{equation}
\begin{equation}
    \mathbf{P}^{(3)}= \begin{pmatrix} 0.45 & 0.1 & 0.3 & 0.35 & 0.25 \\ 0.2 & 0.35 & 0.1 & 0.15 & 0.2 \\ 0.35 & 0.2 & 0.5 & 0.4 & 0.15 \\ 0.3 & 0.4 & 0.25 & 0.55 & 0.35 \\ 0.25 & 0.3 & 0.15 & 0.1 & 0.4  \end{pmatrix} \hspace{5mm}
    \mathbf{P}^{(4)}= \begin{pmatrix} 0.6 & 0.1 & 0.5 & 0.2 & 0.25 \\ 0.2 & 0.65 & 0.15 & 0.5 & 0.1 \\ 0.25 & 0.3 & 0.4 & 0.35 & 0.15 \\ 0.3 & 0.15 & 0.25 & 0.45 & 0.35 \\ 0.4 & 0.2 & 0.3 & 0.15 & 0.55  \end{pmatrix}.
\end{equation}

\newpage
\section{FAO trade data: list of member countries of the 5 clusters}\label{list_of_countries}

\begin{table}[ht]
    \caption{List of member countries of the 5 clusters}
    \centering
    \setlength{\tabcolsep}{2pt}
    \small
    \begin{tabular}{c|c|c|c|c}
        \textcolor{my_orange}{Cluster 1} & \textcolor{my_purple}{Cluster 2} & \textcolor{my_blue}{Cluster 3} & \textcolor{my_red}{Cluster 4} & \textcolor{my_green}{Cluster 5} \\
        \hline
        Denmark & France & Belarus & Afghanistan & New Zealand \\ 
Ireland & Germany & Latvia & Peru & South Africa \\ 
Austria & Italy & Finland & Sri Lanka & Oman \\ 
Poland & Netherlands & Bosnia and Herzegovina & Tanzania & Bahrain \\ 
Sweden & Russia & Bulgaria & Ghana & Jordan \\ 
Switzerland & Spain & Cyprus & Azerbaijan & India \\ 
Turkey & Belgium & Hungary & Guatemala & Indonesia \\ 
Ukraine & UK & Lithuania & Iran & Australia \\ 
Croatia & USA & Luxembourg & Pakistan & Japan \\ 
Czech Republic & China & Serbia & Argentina & Lebanon \\ 
Greece & & Slovakia & Brazil & Malaysia \\ 
Israel & & Slovenia & Cameroon & Philippines \\ 
Norway & & Macedonia & Colombia & South Korea \\ 
Portugal & & Estonia & Costa Rica & Saudi Arabia \\ 
Romania & & & Côte d'Ivoire & Singapore \\ 
& & & Ecuador & Thailand \\ 
& & & Kenya & Canada \\ 
& & & Nicaragua & Mexico \\ 
& & & Syria & Morocco \\ 
& & & Tunisia & \\ 
& & & Chile & \\ 
& & & El Salvador & \\
\end{tabular}
\end{table}

\newpage
\bibliography{refs}
\end{document}